\documentclass[sigconf, nonacm]{acmart}
\usepackage{tikz}
\usepackage{bbding}
\usepackage{cancel}
\usepackage{subcaption}
\usepackage{hyperref}
\usepackage{newtxmath}
\usepackage{enumitem}
\usepackage{amsmath}
\usepackage{pifont}
\usepackage[ruled, lined, linesnumbered, commentsnumbered, noend]{algorithm2e}
\usepackage{tabularx}
\usepackage{graphicx}
\usepackage{textcomp}
\usepackage{caption}
\usepackage{multirow}
\usepackage[normalem]{ulem}
\useunder{\uline}{\ul}{}
\definecolor{edit}{rgb}{0,0,0}
\definecolor{revision}{rgb}{0,0,0}
\newcommand\vldbdoi{XX.XX/XXX.XX}
\newcommand\vldbpages{XXX-XXX}
\newcommand\vldbvolume{18}
\newcommand\vldbissue{1}
\newcommand\vldbyear{2025}
\newcommand\vldbauthors{\authors}
\newcommand\vldbtitle{\shorttitle} 
\newcommand\vldbavailabilityurl{https://github.com/shengyufan/ACE}
\newcommand\vldbpagestyle{plain} 

\newcommand{\tabincell}[2]{\begin{tabular}{@{}#1@{}}#2\end{tabular}}

\newcommand{\name}{ACE\xspace}

\newtheorem{definition}{Definition}

\begin{document}
\title{\name: A Cardinality Estimator for Set-Valued Queries}

%%
%% The "author" command and its associated commands are used to define the authors and their affiliations.
\author{Yufan Sheng}
\affiliation{
    \institution{University of New South Wales}
    \city{Sydney}
    \country{Australia}
}
\email{yufan.sheng@unsw.edu.au}

\author{Xin Cao}
%\authornote{Corresponding author.}
\affiliation{
    \institution{University of New South Wales}
    \city{Sydney}
    \country{Australia}
}
\email{xin.cao@unsw.edu.au}

\author{Kaiqi Zhao}
\affiliation{
    \institution{The University of Auckland}
    \city{Auckland}
    \country{New Zealand}
}
\email{kaiqi.zhao@auckland.ac.nz}

\author{Yixiang Fang}
\affiliation{
    \institution{The Chinese University of Hong Kong, Shenzhen}
    \city{Shenzhen}
    \country{China}
}
\email{fangyixiang@cuhk.edu.cn}

\author{Jianzhong Qi}
\affiliation{
    \institution{The University of Melbourne}
    \city{Melbourne}
    \country{Australia}
}
\email{jianzhong.qi@unimelb.edu.au}

\author{Wenjie Zhang}
\affiliation{
    \institution{University of New South Wales}
    \city{Sydney}
    \country{Australia}
}
\email{wenjie.zhang@unsw.edu.au}

\author{Christian S. Jensen}
\affiliation{
    \institution{Aalborg University}
    \city{Aalborg}
    \country{Denmark}
}
\email{csj@cs.aau.dk}

\begin{abstract}
  Cardinality estimation is a fundamental functionality in database systems. Most existing cardinality estimators focus on handling predicates over numeric or categorical data. They have largely omitted an important data type, set-valued data, which frequently occur in contemporary applications such as information retrieval and recommender systems. 
%In this study, we provide a solution to cardinality estimation over set-valued data. 
%Typical query predicates over set-valued data include \emph{superset},  \emph{subset}, and \emph{overlap}. 
The few existing estimators for such data either favor high-frequency elements or rely on a \emph{partial independence assumption}, which limits their practical applicability. 

We propose ACE, an \uline{A}ttention-based \uline{C}ardinality \uline{E}stimator for estimating the cardinality of queries over set-valued data. We first design a distillation-based data encoder to condense the dataset into a compact matrix.
% present a simple yet effective model to encode a set by aggregating the element information. %Since the complexity of the attention mechanism is proportional to the size of calculated parts, 
% Next, we propose a distillation model to compress an entire dataset and obtain a collection of small-sized embeddings to represent the set-valued data with reduced  computational complexity. 
We then design an attention-based query analyzer to capture correlations among query elements.
% using the attention mechanism. 
To handle variable-sized queries, a pooling module is introduced, followed by a regression model (MLP) to generate final cardinality estimates.
We evaluate \name on three datasets with varying query element distributions, demonstrating that \name outperforms the state-of-the-art competitors in terms of both accuracy and efficiency.
\end{abstract}

\maketitle

\pagestyle{\vldbpagestyle}
\begingroup\small\noindent\raggedright\textbf{PVLDB Reference Format:}\\
\vldbauthors. \vldbtitle. PVLDB, \vldbvolume(\vldbissue): \vldbpages, \vldbyear.\\
\href{https://doi.org/\vldbdoi}{doi:\vldbdoi}
\endgroup
\begingroup
\renewcommand\thefootnote{}\footnote{\noindent
This work is licensed under the Creative Commons BY-NC-ND 4.0 International License. Visit \url{https://creativecommons.org/licenses/by-nc-nd/4.0/} to view a copy of this license. For any use beyond those covered by this license, obtain permission by emailing \href{mailto:info@vldb.org}{info@vldb.org}. Copyright is held by the owner/author(s). Publication rights licensed to the VLDB Endowment. \\
\raggedright Proceedings of the VLDB Endowment, Vol. \vldbvolume, No. \vldbissue\ %
ISSN 2150-8097. \\
\href{https://doi.org/\vldbdoi}{doi:\vldbdoi} \\
}\addtocounter{footnote}{-1}\endgroup

\ifdefempty{\vldbavailabilityurl}{}{
\vspace{.3cm}
\begingroup\small\noindent\raggedright\textbf{PVLDB Artifact Availability:}\\
The source code, data, and/or other artifacts have been made available at \url{\vldbavailabilityurl}.
\endgroup
}

\section{Introduction}
\label{sec1}
\textcolor{revision}{Set-valued data where the value of an attribute is a set of elements has emerged as an essential data type in many real-world applications, including information retrieval~\cite{castro12023information}, %machine learning~\cite{wang2023set}, 
recommender systems~\cite{altosaar2021rankfromsets}, and social networks~\cite{lachlan2016social}. 
For example, in a movie recommender system, each movie's genre is generally associated with a set of categories such as sci-fi, action, and comedy. 
In X (\url{http://www.twitter.com}), each tweet generally has multiple hashtags. Table~\ref{tab:example} shows a toy example, where each row corresponds to a tweet and the \textit{Hashtags} column is a set-valued attribute that stores the hashtags of a tweet.}

\textcolor{revision}{Given the prevalence of set-valued data across different domains, set queries play a crucial role in efficiently handling multi-valued attributes and complex relationships. 
%
%Unlike traditional queries that operate on single values, set queries enable operations such as subset, superset, and intersection, which are essential for handling complex relationships. 
For example, X offers the functionality of searching for tweets by a set of keywords or hashtags using operators such as ``OR'' and ``AND.'' The SQL standard includes support for storing multi-valued data in a single row~\cite{kleppmann2017designing}. Set-valued data and set queries are supported to varying degrees by modern DBMSs such as Oracle~\cite{oracle23array}, MySQL~\cite{mysql17set}, IBM DB2~\cite{ibm2024db2}, SQL Server~\cite{sqlserver23tab}, and PostgreSQL~\cite{pg2024}. For example, MySQL supports up to 64 distinct elements in a set-valued attribute~\cite{mysql17set}. SQL Server enables passing a set as a table-valued parameter. To the best of our knowledge, PostgreSQL offers the best support for set-valued data and set queries. It provides three set query predicates: \emph{superset} (\texttt{@>}), \emph{subset} (\texttt{<@}), and \emph{overlap} (\texttt{\&\&}).}
For example, to evaluate public interest in the recent United States presidential election, we can count the number of tweets containing at least one presidential candidate. This can be achieved using the following set query $q$ in PostgreSQL: \texttt{SELECT COUNT(*) FROM \textit{T} WHERE \textit{T}.Hashtags \&\& ARRAY["Trump", "Harris"].
}

%, and it is the only RDBMS that offers a built-in estimator for set operators to our best knowledge.
% For applicability, we focus on the three set query predicates supported in PostgreSQL: \emph{superset} (\texttt{@>}), \emph{subset} (\texttt{<@}), and \emph{overlap} (\texttt{\&\&}). 

%In social networks, each tweet posted can be tagged with multiple hashtags, and Twitter offers the function to search for tweets by a set of keywords or hashtags. Table~\ref{tab:example} shows a toy example on X (\url{http://www.twitter.com}). Each row in the table corresponds to a posted tweet, and the \textit{Hashtags} column is a set-valued attribute that stores the hashtags of each tweet. }
%

%To evaluate public interest in the recent United States presidential election, we can count the number of tweets containing at least one presidential candidate. This can be achieved using the following set query $q$: \texttt{SELECT COUNT(*) FROM \textit{T} WHERE \textit{T}.Hashtags \&\& ARRAY["Trump", "Harris"]}.}

\begin{table}[tb]
\small
\caption{Twitter hashtag dataset}
\vspace{-0.4cm}
\begin{tabular}{|c|c|}
\hline
\textbf{Tweet\_ID} & \textbf{Hashtags}                     \\ \hline
$t_{1}$              & \{Trump, shot\}                      \\ \hline
$t_{2}$              & \{Spain, Euros, Yamal\}              \\ \hline
$t_{3}$              & \{Biden, Harris, Trump\}            \\ \hline
$t_{4}$              & \{Harris, Trump, debate\}             \\ \hline
$t_{5}$              & \{JD Vance, Trump\}                  \\ \hline
$t_{6}$              & \{Messi, Yamal\}                     \\ \hline
$t_{7}$              & \{Messi, Argentina, Copa America\} \\ \hline
\end{tabular}
\label{tab:example}
\end{table}

%\textcolor{revision}{%Traditional database management systems (DBMSs) are primarily based on the relational model that emphasizes adherence to the first normal form~\cite{codd1970relational}.
%With the prevalence of set-valued data and increasing demands for handling, the SQL standard adds support, which allows multi-valued data to be stored within a single row~\cite{kleppmann2017designing}. This feature is also supported to varying degrees by Oracle~\cite{oracle23array}, MySQL~\cite{mysql17set}, IBM DB2~\cite{ibm2024db2}, SQL Server~\cite{sqlserver23tab}, and PostgreSQL~\cite{pg2024}.}
% current DBMSs provide only limited support for such data.
% MySQL supports the set data type but only up to 64 distinct elements in a set-valued attribute value~\cite{mysql17set}. Although SQL Server enables passing a set in a table-valued parameter, it does not support cardinality estimation over set-valued data~\cite{sqlserver23tab}. PostgreSQL includes an array data type for set-valued data, and it is the only RDBMS that offers a built-in estimator for set operators to our best knowledge.
% For applicability, we focus on the three set query predicates supported in PostgreSQL: \emph{superset} (\texttt{@>}), \emph{subset} (\texttt{<@}), and \emph{overlap} (\texttt{\&\&}).}

To identify efficient query execution plans for complex queries, cardinality estimation of a query step plays a crucial role since it 
directly influences the efficiency of database query execution. 
%
%The goal of cost-based query optimization is to identify efficient query execution plans by estimating the costs of query plans. Cardinality estimation of a query step plays a crucial role in the process because it directly affects the accuracy of the overall cost estimates.
%
%the result size or cardinality of a query or sub-query is used to assign a cost to each enumerated query execution plan, which is used in turn to find the plan with the lowest overall cost. Since actual cardinalities are unknown before executing queries, a typical query optimizer estimates them by multiplying the size of the input with the estimated selectivity~\cite{selinger1979access}. 
Cardinality estimation has been extensively studied~\cite{hasan2020deep, ioannidis1991propagation, ioannidis2003history, li2016wander, poosala1997selectivity, wu2021unified}, showing its profound impact on the quality of selected query plans~\cite{leis2015good, han2021cardinality}. 
However, most DBMSs provide only limited support for optimizing set query execution. %For example, SQL Server enables passing a set as a table-valued parameter, but it cannot perform cardinality estimation for set-valued predicates. 
To our knowledge, PostgreSQL is the only DBMS that offers a built-in estimator for set operators but the accuracy is not good enough.  
In this study, we investigate cardinality estimation for queries over set-valued data, which has not received sufficient attention.
While some cardinality estimators for set-valued predicates exist~\cite{yang2019selectivity,meng2023selectivity,korotkov2016selectivity}, they each have significant shortcomings.

First, most studies \textbf{pay more attention to elements with high frequency (P1).} For example, Yang et al.~\cite{yang2019selectivity} propose two sampling-based cardinality estimators for subset queries that aim to capture the distribution of high-frequency elements. They suffer in accuracy over queries containing low-frequency elements~\cite{meng2023selectivity}.

Second, most existing estimators \textbf{do not capture the correlation among elements in a query well (P2)}, which is crucial for accurate estimation. For example, ``Harris'' and ``Trump'' appear 2 and 4 times, respectively, in the example. If we assume independence, the estimated cardinality for $q$ is $2+4=6$, while the actual cardinality is 4 because $t_{3}$ and $t_{4}$ contain both keywords. %In general, the element correlations influence cardinality estimation markedly. 
%, highlighting the need to account for element correlations during estimation. %Nevertheless, PostgreSQL regards each element as a separate attribute and introduces a basic model for cardinality estimation of set-valued queries. 
Korotkov et al.~\cite{korotkov2016selectivity} leverage a probabilistic model~\cite{getoor2001selectivity} to address the element correlation issue. However, their model still relies on random sampling of high-frequency elements, thus missing the correlation for low-frequency elements. Recently, Meng et al.~\cite{meng2023selectivity} propose to convert a set-valued column into multiple categorical columns. They then utilize existing estimators to capture the correlation between columns. This approach still ignores the correlation among elements within the same subcolumn, leading to unstable estimation accuracy. Besides, this solution relies on that a set query can be converted into categorical sub-queries, which does not support all set queries such as the overlap query.
%
% \begin{table}[tb]
% \caption{Set Transformer as the cardinality estimator}
% \vspace{-0.3cm}
% \label{tab:pre}
% \resizebox{\columnwidth}{!}{%
% \begin{tabular}{|c|cccc|cccc|cccc|}
% \hline
% \multirow{2}{*}{\begin{tabular}[c]{@{}c@{}}Query\\ Type\end{tabular}} & \multicolumn{4}{c|}{Regular} & \multicolumn{4}{c|}{High-frequency} & \multicolumn{4}{c|}{Low-frequency} \\ \cline{2-13} 
%                                                                       & Mean  & 50\%  & 95\%  & 99\% & Mean    & 50\%    & 95\%   & 99\%   & Mean    & 50\%   & 95\%   & 99\%   \\ \hline
% Superset                                                              & 21.4  & 8.92  & 75.1  & 119  & 14.8    & 5.47    & 60.7   & 98.7   & 37.5    & 32.7   & 85.8   & 106    \\ \hline
% Subset                                                                & 2.42  & 1.64  & 6.45  & 14.8 & 3.04    & 2.07    & 7.91   & 12.2   & 7.79    & 5.39   & 19.8   & 37.2   \\ \hline
% Overlap                                                               & 156   & 65.2  & 706   & 1087 & 249     & 115     & 887    & 1652   & 12.6    & 4.08   & 45.2   & 181    \\ \hline
% \end{tabular}%
% }
% \end{table}

Third, existing studies mainly focus on capturing data distribution and \textbf{overlook the valuable insights in historical query workloads (P3).} The cardinalities of two similar queries can differ on the operators used, even when their elements are identical. For example, the cardinality of the example query ($q = T$ \texttt{\&\&} \{"Harris", "Trump"\}) is 4 while the cardinality of a similar superset query ($q' = T$ \texttt{@>} \{"Harris", "Trump"\}) is 2. Learning the data distribution only is insufficient for accurate predictions across various query types. Set Transformer~\cite{lee2019set} processes input sets using an attention mechanism to capture the correlations between elements, making it a potential candidate for a query-driven estimator. However, our experiments in Section~\ref{sec7} show that its accuracy is unstable and sometimes performs much worse than data-driven estimators, because it is impractical to represent all possible combinations of elements given limited training data. Thus, pure query-driven methods that treat the problem as a supervised learning task also have a severe issue: their accuracy highly depends on the quantity and quality of the training data (i.e., known query workload)~\cite{han2021cardinality}.
% As demonstrated in the previous work~\cite{li2023alece}, utilizing information from the query workload can improve the estimator's performance. However, pure query-driven methods have their inherent limitations, that is, the estimation performance heavily depends on the quantity and quality of training data.

\begin{figure}[!htb]    
    \centering
    \includegraphics[width=.9\linewidth]{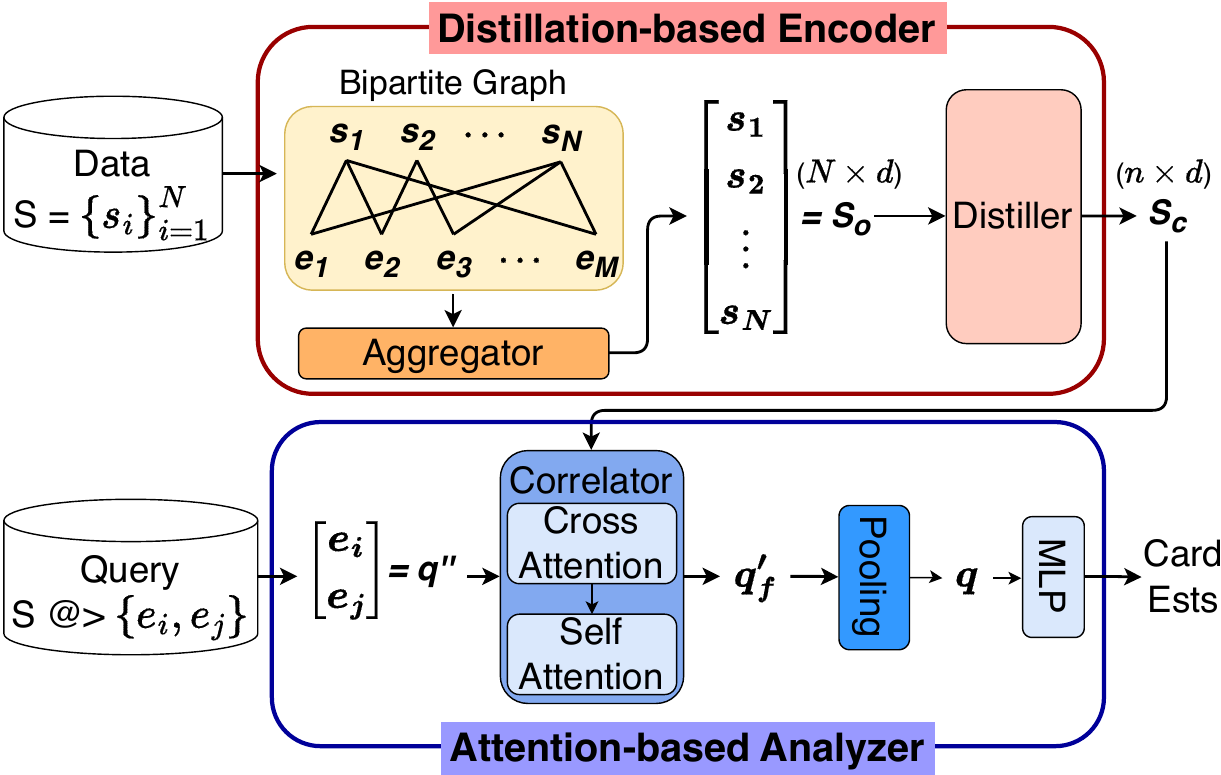}
    %\vspace{-0.3cm}
    \caption{Overview of \name.}
    \label{fig:view}
\end{figure}

To address the issues above, we propose ACE, an \uline{A}ttention-based \uline{C}ardinality \uline{E}stimator for queries over set-valued data. As depicted in Figure~\ref{fig:view}, \name leverages information from both the data and the query workload to address \textbf{P3}.

To address \textbf{P1}, we design a distillation-based data encoder to generate a compact dataset representation. We construct a bipartite graph that models the relationships between the set elements and their corresponding sets. This graph serves as the foundation for the subsequent aggregation step. The aggregator module synthesizes a set embedding by integrating information from all elements of the set, ensuring that even low-frequency elements are not underrepresented. Once the set embeddings are computed, they are concatenated to form the initial representation of the full dataset (i.e., a database table of set-valued data), denoted as $\boldsymbol{S_{o}}$. For large datasets, this representation has a high dimensionality, posing significant challenges for downstream learning tasks. To address this issue, a distiller module is designed to produce a more compact representation, $\boldsymbol{S_{c}}$, which compresses the original representation with a fixed ratio while preserving as much information as possible.

%Motivated by existing studies~\cite{vaswani2017attention, zhao2022queryformer, li2023alece}, 
Next, we design a correlator module to capture element correlations and address \textbf{P2}. For each query in a given workload, 
%\textcolor{edit}{
we apply a cross-attention mechanism to generate the query element embeddings using the data representation obtained from the previous step. 
The computational complexity of the attention mechanism~\cite{bahdanau2014neural, vaswani2017attention, choromanski2020rethinking, shen2021efficient} scales with the size of the input, i.e., the number of rows in the data representation and the number of query elements, which emphasizes the necessity of the distillation step. Then, we utilize the self-attention mechanism to capture the correlation between the learned latent representation of query elements. It is noteworthy that set-valued queries have varying sizes, bringing another challenge for the learning-based estimator. We address the challenge with a pooling module to generate a fixed-sized query embedding $\boldsymbol{q}$ and finally a linear regression model to map the embedding to a cardinality estimation.

\begin{table}[tb]
\caption{Properties of different estimators}
\vspace{-0.3cm}
\label{tab:property}
\resizebox{\columnwidth}{!}{%
\begin{tabular}{|c|ccc|c|c|c|c|}
\hline
\multirow{2}{*}{Method} & \multicolumn{3}{c|}{Query supported}         & 
        %\multirow{2}{*}{Data-driven}
        \multirow{2}{*}{\tabincell{c}{Data-\\driven}}
   & 
        %\multirow{2}{*}{Query-driven}
        \multirow{2}{*}{\tabincell{c}{Query-\\driven}}
   & \multirow{2}{*}{\begin{tabular}[c]{@{}c@{}}Low-frequency\\ elements\end{tabular}} & \multirow{2}{*}{\begin{tabular}[c]{@{}c@{}}Element \\correlation\end{tabular}} \\ \cline{2-4}
                        & Superset     & Subset       & Overlap      &                              &                               &                                        &                                                                                            \\ \hline
PostgreSQL              & $\checkmark$ & $\checkmark$ & $\checkmark$ & $\checkmark$                 & $\times$                      & $\times$                               & $\times$\\
Sampling            & $\checkmark$ & $\checkmark$ & $\checkmark$ & $\checkmark$                 & $\times$                      & $\times$                               & $\times$\\
OT-S~\cite{yang2019selectivity}             & $\checkmark$ & $\checkmark$ & $\checkmark$ & $\checkmark$                 & $\times$                      & $\times$                               & $\times$\\
ST~\cite{meng2023selectivity}                      & $\checkmark$ & $\checkmark$ & $\times$     & $\checkmark$                 & $\times$                      & $\checkmark$                           & $\bcancel{\checkmark}$\\
STH~\cite{meng2023selectivity}                     & $\checkmark$ & $\checkmark$ & $\times$     & $\checkmark$                 & $\times$                      & $\checkmark$                           & 
$\bcancel{\checkmark}$\\
\name(Ours)                   & $\checkmark$ & $\checkmark$ & $\checkmark$ & $\checkmark$                 & $\checkmark$                  & $\checkmark$                           & $\checkmark$                                                                                    \\ \hline
\end{tabular}%
}
\end{table}

Table~\ref{tab:property} summarizes the novelty of our estimator \name compared with existing set-valued query cardinality estimators. \textcolor{revision}{Please note that we use the symbol $\bcancel{\checkmark}$ because ST and STH can only capture correlations between elements in different columns.} Overall, we make the following contributions:
\begin{itemize}
    \item We propose \name, a learning-based cardinality estimator for queries over set-valued data, exploiting both the data and query workload distributions.
    \item We design a distillation-based data encoder to generate a dataset compact representation, reducing the dimensionality while retaining key information.
    % for accurate cardinality estimation.
    \item We propose an attention-based query analyzer that 
    % links query elements with underlying data and 
    captures correlations among query elements, followed by a pooling method to address the issue of variable-sized queries.
    \item We compare \name and the state-of-the-art estimators on real-world datasets and query workload. The results show that \name outperforms the SOTA estimators by up to 33.9$\times$ in terms of accuracy while offering a stable estimation latency.
    \textcolor{revision}{Additionally, the integration of \name and PostgreSQL also speeds up the end-to-end execution for complex queries.}
\end{itemize}

% {\color{red}Can omit the paragraph to save space.}
%The rest of the paper is organized as follows. Section~\ref{sec2} defines the problem and briefly describes the attention mechanism. Section~\ref{sec3} presents the overall framework. Sections~\ref{sec4} and~\ref{sec5} elaborate on the dataset encoder and the query analyzer. We explain the workflow for handling dynamic data in Section~\ref{sec6}. Sections~\ref{sec7} and~\ref{sec8} report on the experimental study and review related work, respectively. Section~\ref{sec9} concludes the paper.

\section{Preliminaries}
\label{sec2}
We start with our problem statement and technical background for our proposed model. 
Table~\ref{tab:notas} lists the frequently used notation.

\begin{table}[htb]
\caption{Frequently used notation}
\vspace{-0.3cm}
\label{tab:notas}
\resizebox{\linewidth}{!}
{%
\begin{tabular}{c|l}
\hline
\textbf{Notation} & \textbf{Meaning} \\ \hline
$S = \{s_i\}_{i=1}^{N}$ & The set-valued dataset                             \\ \hline
$E = \{e_j\}_{j=1}^{M}$ & The element universe of the dataset                \\ \hline
$d$                     & The dimension of the embedding                     \\ \hline
$B_d$/$B_q$             & The batch size of data/query                     \\ \hline
$r$        & The distillation ratio                     \\ \hline
$n_\mathit{distill}$        & The number of layers for the distillation model                     \\ \hline
$n_\mathit{cross}$/$n_\mathit{self}$             & The number of layers for the cross/self attention                     \\ \hline
$\boldsymbol{s}$/$\boldsymbol{e}$/$\boldsymbol{q}$   & The set/element/query embedding                          \\ \hline
$\boldsymbol{S_o}$/$\boldsymbol{S_c}$ & The original/distilled matrix of the dataset       \\ \hline
$\boldsymbol{Q}$/$\boldsymbol{K}$/$\boldsymbol{V}$ & The queries/keys/values of the attention mechanism \\ \hline
\end{tabular}
}
\end{table}

\subsection{Problem Statement}
\begin{definition}[\textbf{Set-Valued Query}]
    A set-valued query $q= (\mathit{operator}, \mathit{literal})$ is a predicate over the set-valued data, represented by an operator-literal pair. To be consistent with PostgreSQL, operator can be the superset (\texttt{@>}), subset (\texttt{<@}), or overlap (\texttt{\&\&})\footnote{PostgreSQL Array Functions and Operators: \href{https://www.postgresql.org/docs/9.1/functions-array.html}{https://www.postgresql.org/docs/9.1/ functions-array.html}} operator, while literal is a subset of $E$. 
\end{definition}

For example, $q = S$ \texttt{@>} $\{e_1, e_3, e_7\}$ is to find the sets over $S$, each of which is a superset of $\{e_1, e_3, e_7\}$.

\noindent \textbf{Problem}. The study aims to propose an estimator that can accurately and efficiently predict the cardinality of a set query without executing the query.

\subsection{Attention Mechanism}
\label{sec2.2}
The attention mechanism was originally envisioned as an enhancement to the encoder-decoder Recurrent Neural Network (RNN) in sequence-to-sequence applications~\cite{bahdanau2014neural}. In neural networks, attention is a technique that aims to mimic human cognitive attention, and its motivation is that a network should focus on the important parts of the data rather than treating all data equally. It employs an attention function to decide which part of the data should be emphasized. This function maps a query and a collection of key-value pairs, assigns weights by computing the similarity between each pair of the query and a key using some metric, and calculates the weighted sum of values as its output. Therefore, compared to other neural networks, the attention mechanism can achieve better interpretability and have higher representative abilities.

In this study, we use the standard Scaled Dot-Product Attention~\cite{vaswani2017attention}, called \textit{Att}. The keys and values are mapped to matrices $\boldsymbol{K}$ and $\boldsymbol{V}$, of dimensions $n \times d_k$ and $n \times d_v$, where $n$, $d_k$, and $d_v$ denote the size of the original input and the dimensions of the matrices $\boldsymbol{K}$ and $\boldsymbol{V}$. A query is first converted to an $m \times d_k$ matrix $\boldsymbol{Q}$, where $m$ indicates the size of the query and is used as input to the dot product. 
If $\boldsymbol{Q}$ is different from $\boldsymbol{K}$ and $\boldsymbol{V}$, it is called cross-attention. Otherwise, it is self-attention. 
Since a large dot product result often leads to the vanishing gradient problem, the \textit{Att} function divides the dot product by the factor $\sqrt{d_k}$. The process consists of calculating the dot products of $\boldsymbol{Q}$ with all keys $\boldsymbol{K}$, dividing each by $\sqrt{d_k}$, applying the softmax function to obtain the weights, and obtaining the output by multiplying the weights and the values $\boldsymbol{V}$.
\begin{equation*}
    \mathit{Att(\boldsymbol{Q}, \boldsymbol{K}, \boldsymbol{V}) = softmax\left ( \frac{\boldsymbol{Q} \cdot \boldsymbol{K}^T}{\sqrt{d_k}} \right )\boldsymbol{V}}
\end{equation*}

Next, we adopt the multi-head attention mechanism~\cite{vaswani2017attention}, which linearly projects the queries, keys, and values using $h$ different linear projections and then computes the \textit{Att} function in parallel. The independent attention outputs are then concatenated and linearly transformed into the expected dimension. Compared with single-head attention, this approach processes different projected spaces jointly, thus capturing complex patterns from different perspectives.
% \begin{equation*}
%     \begin{aligned}
%         MultiHead(\boldsymbol{Q}, \boldsymbol{K}, \boldsymbol{V}) &= concat(head_1, \cdots, head_h) W_o, \\
%         \mbox{where } head_i &= Att(\boldsymbol{Q}W_{i}^{Q}, \boldsymbol{K}W_{i}^{K}, \boldsymbol{V}W_{i}^{V})
%     \end{aligned}
% \end{equation*}
%
\begin{equation*}
    \mathit{MultiHead(\boldsymbol{Q}, \boldsymbol{K}, \boldsymbol{V}) = concat(head_1, \cdots, head_h) W_o},
\end{equation*}
where $\mathit{head_i = Att(\boldsymbol{Q}W_{i}^{Q}, \boldsymbol{K}W_{i}^{K}, \boldsymbol{V}W_{i}^{V})}$.

\section{Overview of \name}
\label{sec3}
The structure of \name is shown in Figure~\ref{fig:view}. The encoder (Section~\ref{sec4}) generates a compact data representation, while the analyzer (Section~\ref{sec5}) captures correlations between query elements by learning from both the queries and the underlying data.
% The cardinality estimates eventually produced in \name depend on both data and queries.

As in previous studies~\cite{deepdb, li2023alece}, the first task is representing the dataset properly. Sets in \textit{S} are combinations of elements, and we can naturally represent a set as the concatenation of the embeddings of its elements.
However, this representation is incompatible with neural networks due to the variable sizes of sets. We need to convert variable-sized sets into fixed-sized vectors. Traditional methods, including padding and truncation, have limitations. For example, padding causes storage overheads and increases time complexities. Instead, we propose to learn the representations of sets.
Assuming the number of elements that can occur in sets is $M$, there are $2^M - 1$ possible sets, making it hard to design one model to represent all the sets. We propose to construct a bipartite graph to model the dependencies between elements and sets and to learn an aggregator to obtain a set embedding $\boldsymbol{s}$ by aggregating the information of each element $\boldsymbol{e}$ in the set. Thus, the underlying dataset is represented by a matrix $\boldsymbol{S_o}$ where each row is the embedding of a set.

Next, we aim to learn the representation of the query elements from data. 
%Existing methods fail to leverage historical queries to effectively link queries with the underlying data. 
This motivates us to employ a cross-attention mechanism to discover the relation between the underlying dataset and each query element. The original data matrix $\boldsymbol{S_o}$ cannot be used directly in the attention framework for large-scale datasets. For instance, training on our smallest dataset \textbf{GN} requires 42GB of GPU memory, even with a batch size of 1. Thus, we design a distiller module to obtain a matrix $\boldsymbol{S_c}$ that preserves the essential knowledge in $\boldsymbol{S_{o}}$.

As shown in the example query in Section~\ref{sec1}, it is crucial to capture correlations between query elements, influenced by the underlying data, to get accurate estimates. Thus, we propose a correlator module to achieve this. We first leverage a data-query cross-attention to measure the relevance between each query element and the underlying data. We then adopt a self-attention mechanism to capture the correlations between the query elements. In addition, the attention mechanism is suitable for dealing with sets as the order of set elements does not affect the output.

To handle the variable-size queries, we utilize a pooling module to derive a fixed-sized vector. This vector extracts pertinent information from the output of the self-attention mechanism and adapts its focus based on the operator type of a set-valued query.

% \noindent \textbf{Workflow.}
% \textcolor{edit}{Given a set-valued dataset $S = \{s_i\}_{i=1}^{N}$, our framework begins by applying an aggregator to generate embeddings for each set in the dataset. The resulting embeddings are concatenated to construct a dataset representation with dimensions $N \times d$. However, for large datasets, this representation becomes impractical for use in the query analyzer due to its size. To address this issue, we introduce a distiller module that compresses the embeddings into a more compact matrix of dimensions $n \times d$ (where $n \ll N)$. Notably, this distillation process is performed independently of the query and can be precomputed once the dataset is available.}

% \textcolor{edit}{When seeing a set-valued query $q$, we apply a cross-attention mechanism to encode the interaction between the query elements and the dataset. This step results in a latent representation of the query elements, which is subsequently processed using a self-attention mechanism to capture intra-query correlations. As the output of the self-attention mechanism has dimensions $k \times d$, where $k$ denotes the number of query elements (varying across different queries), we utilize a pooling module to generate a fixed-sized vector. Finally, we use a simple linear regression model to predict the cardinality based on this vector.}

\noindent \textbf{Offline training.} The training process is divided into two distinct phases. In the first phase,  we employ an unsupervised learning approach to train the data encoder, requiring only a small subset of the dataset. In the second phase, we utilize a supervised learning method to train the query analyzer, using both the query embeddings and the distilled matrix generated by the encoder as input. During this stage, true query cardinalities serve as ground-truth labels. The entire training procedure leverages stochastic gradient descent (SGD) optimization~\cite{bottou-98x}.

\noindent \textbf{Online estimation.} In the pre-processing phase, a well-trained data encoder can distill the entire dataset into a compact data matrix. When a new query arrives, we can only utilize the learned query analyzer to estimate the cardinality efficiently, taking the query element embeddings and the matrix as input.

\section{Dataset Featurization}
\label{sec4}
When representing the set-valued dataset, PostgreSQL uses histograms to approximate the distribution of the underlying data. A recent study~\cite{meng2023selectivity} converts a set into a smaller number of numerical values, models the factorization problem as a graph coloring problem, and proposes a greedy method to address the NP-hard problem. However, this method cannot measure the correlation between the elements in the same partition. Recently, machine learning techniques have opened the opportunity to learn models that outperform many traditional methods~\cite{zhang2024pace, kraska2018case}. Thus, we aim to learn a model that encodes each set and generates the data representation. We also design a distillation model such that the featurization matrix can be effectively used in the attention mechanism. The details are given in Sections~\ref{sec4.1} and~\ref{sec4.2}.

\subsection{Set Representation}
\label{sec4.1}
\begin{figure}[htb]
    \centering
    \subcaptionbox{Element graph\label{fig:graph1}}{
        \vspace{-0.2cm}
        \includegraphics[width=.475\linewidth]{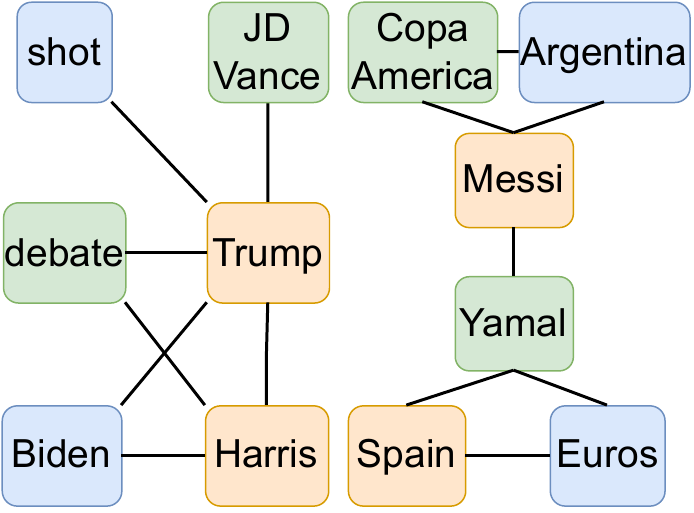}
    }
    \subcaptionbox{Element-set graph\label{fig:graph2}}{
        \vspace{-0.2cm}
        \includegraphics[width=.475\linewidth]{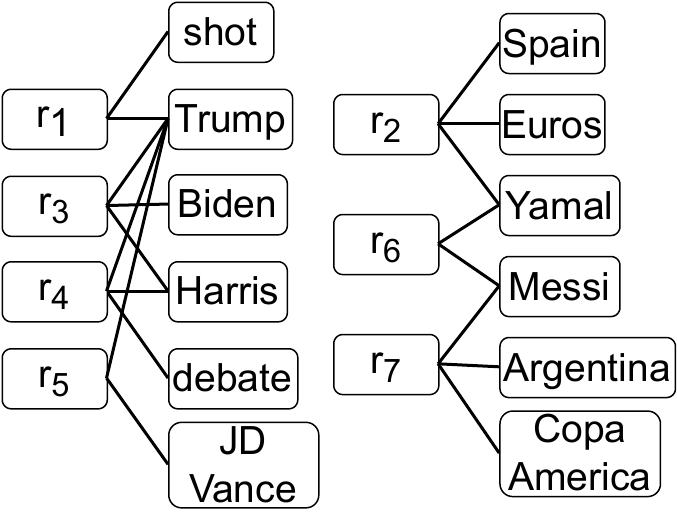}
    }
    \vspace{-0.3cm}
    \caption{Graph construction approaches.}
    \Description{Two methods for constructing the graph}
\end{figure}

We follow the setting used in existing studies~\cite{lee2022set, korotkov2016selectivity} where the element universe $E$ is finite and fixed, meaning each set consists of known elements. To partition a set into several clusters, the existing work~\cite{meng2023selectivity} builds an undirected graph based on the underlying data, where edges connect two elements that appear in the same set and uses a greedy algorithm to partition elements into $k$ clusters. Taking the scenario of $k = 3$ as an example, the algorithm proceeds in two phases. In the first stage, it builds a graph and uses the largest first algorithm~\cite{kosowski2004classical} to obtain initial partitions, ensuring that no elements within a partition are contained in the same set. In the second stage, the algorithm greedily merges partitions to produce the result clusters, as illustrated in Figure~\ref{fig:graph1}, where elements of the same color belong to the same cluster. Although they utilize the existing works~\cite{deepdb, neurocard} to capture the correlation among clusters, the correlation between the elements within the same cluster, such as "Trump" and "Harris", cannot be measured. Unlike the previous method, we represent the dataset as embeddings so that we can leverage the machine learning method to capture the correlation between elements appeared in a query.

However, proposing such a model is non-trivial. As analyzed in Section~\ref{sec3}, the number of combinations of $M$ elements equals $2^M - 1$, making it challenging to learn a model that considers all possibilities. Motivated by existing works~\cite{wu2023billion, chen2023bipartite, zhang2023contrastive}, we build a bipartite graph to model the correlation between elements and sets, as shown in Figure~\ref{fig:graph2}. In this graph, an edge connects an element to a set if the element appears in that set. Now, the problem shifts to representing a set $s$ when its comprised elements are known. 

Following this motivation, we propose an aggregator module that takes element embeddings as input.
\textcolor{revision}{A na\"ive approach is to adopt pooling methods directly. However, pooling methods discard considerable amounts of information. For example, max-pooling only retains the highest value, ignoring the rest. Additionally, pooling methods are non-trainable operations that cannot adapt to various data, limiting their ability to extract complex representations and potentially leading to suboptimal feature compression.
Prior studies~\cite{ying2018graph, hamilton2017inductive, wang2018deep, zhang2019deep} show that the Multi-Layer Perceptron (MLP)~\cite{hastie2009elements} offers a simple yet effective approach to compute feature representations for each input. Thus, we utilize an MLP to aggregate the information from elements before performing pooling.} 
% However, using an MLP presents a challenge: it cannot create a fixed-size embedding when the input size varies. To tackle this issue, we incorporate a pooling operator. 
The process of generating the set embedding $\boldsymbol{s}$ is described as follows (where $\boldsymbol{e_j}$ is the embedding of the element $e_j$):
\begin{equation*}
    \mathit{\boldsymbol{s} = Pool \left ( \left \{ MLP\left ( \boldsymbol{e_j} \right ), \forall e_j \in s \right \} \right )}
\end{equation*}

Given the lack of inherent order among elements, any symmetric vector function can be used as the pooling operator. Following the prior work~\cite{hamilton2017inductive}, we utilize the simple single-layer architecture with the mean-pooling operator.

\subsection{Dataset Distillation}
\label{sec4.2}
The above aggregation model can encode each set as a $1 \times d$ embedding and we can concatenate them together to obtain an $N \times d$ matrix $\boldsymbol{S_o}$ representing the dataset, where $N$ denotes the number of sets in the dataset. Motivated by the existing work~\cite{li2023alece}, we aim to use the attention mechanism to link query elements with the dataset representation. However, when dealing with large-scale datasets, directly using $\boldsymbol{S_o}$ is unrealistic. As introduced in Section~\ref{sec2.2}, the correlation in the attention mechanism is captured by the dot product of $\boldsymbol{Q}$ and $\boldsymbol{K^T}$, meaning that the complexity of the mechanism is proportional to the size of the dataset matrix. Since the size of real-world dataset is usually larger than $10^{6}$, we aim to synthesize a small dataset such that models trained on it achieve high performance on the original large dataset.

A na\"ive method is to draw a small sample from the original data matrix. However, the resultant matrix is lossy, and the performance depends heavily on the sampling quality. Recently, the problem of dataset distillation has been studied in the field of computer vision. Existing works~\cite{wang2018dataset, li2020soft, zhao2021DC} introduce different algorithms that take as input a large real dataset to be distilled and output a small synthetic distilled dataset, which is evaluated via testing models trained on this distilled dataset on a separate real dataset. However, these methods cannot be adopted to solve our problem because the dataset studied in previous works always has the label information and they distill the data of the same class into a small dataset. For example, the image dataset can be represented as $T = {\{ \left( x_g, y_g \right)\}}_{g=1}^{G}$ where $G$ denotes the number of training images, $x_g$ and $y_g$ denote the image and its corresponding label, respectively. Then, they propose various approaches that can compress thousands of training images into just several synthetic distilled images (e.g. one per class) and achieve comparable performance with training on the original dataset. However, our dataset representation lacks the essential label information. Therefore, we aim to propose a distillation model that can compress the large unlabeled dataset.

\begin{figure}[htb]
    \centering
    \includegraphics[width=.95\linewidth]{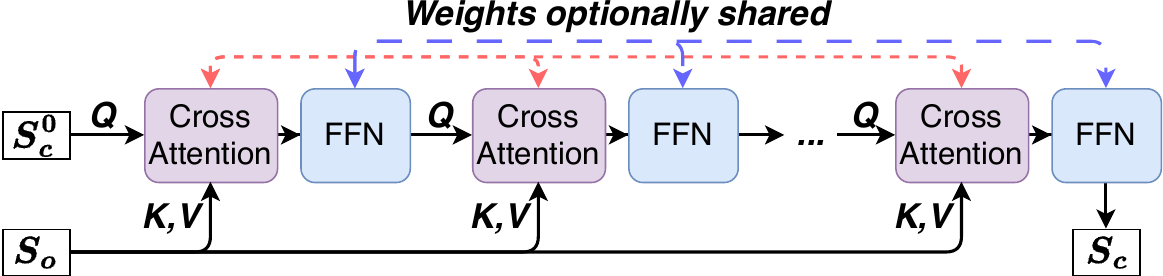}
    \vspace{-0.3cm}
    \caption{Distillation model.}
    \Description{Attention-based distillation model for each batch of data}
    \label{fig:dis}
\end{figure}

As shown in the previous work~\cite{vaswani2017attention}, the self-attention mechanism allows the model to aggregate information across tokens. Building on this, we aim to utilize the attention mechanism to compress the dataset $\boldsymbol{S_o}$. Since self-attention produces an output matrix with the same dimensions as its input, we propose an iterative cross-attention model, as illustrated in Figure~\ref{fig:dis}. 

Each cross-attention block consists of a single attention layer $Att$, followed by a feed-forward neural network $FFN$. Initially, we sample a set of embeddings as the initial value $\boldsymbol{S_{c}^{0}}$. Then, we project the distilled matrix to the query $\boldsymbol{Q}$ while mapping the original matrix to the key $\boldsymbol{K}$ and the value $\boldsymbol{V}$. Note that we adopt residual connections~\cite{he2016deep} and layer normalization~\cite{ba2016layer} in our framework.
\begin{equation*}
    \begin{aligned}
        \boldsymbol{\tilde{S}_c^{i}} &= \mathit{LayerNorm}\left ( \boldsymbol{S_c^{\mathit{i-1}}} + \mathit{Att}\left ( \boldsymbol{S_c^{\mathit{i-1}}}, \boldsymbol{S_o}, \boldsymbol{S_o} \right ) \right ), \\
        \boldsymbol{S_c^{i}} &= \mathit{LayerNorm}\left ( \boldsymbol{\tilde{S}_c^{i}} + \mathit{FFN} \left(\boldsymbol{\tilde{S}_c^{i}} \right) \right )
    \end{aligned}
\end{equation*}

By iteratively applying the cross-attention mechanism, our model can extract useful information from the original matrix while reducing the size of the matrix simultaneously. This model can also be seen as performing the clustering of the inputs with the latent positions as cluster centers, leveraging highly asymmetric cross-attention layers. Following previous works~\cite{xiao2019sharing, jaegle2021perceiver}, we share weights between each instance of the cross-attention module (except the first one) for parameter efficiency. Consequently, we utilize the smaller $\boldsymbol{S_c}$ as input of the following query analyzer.

\subsection{Encoder Training}
\label{sec4.3}
The dataset encoder comprises two distinct modules, each with an optimization objective. To address this, we propose a combined loss function that integrates both objectives, allowing training the two modules simultaneously, in line with previous works~\cite{silver2016mastering, liu2019multi, hao2021ks}.

The aggregation module aims to generate the set embeddings by integrating the information from elements. To achieve this, we predict whether there is an edge connecting the set and the element based on their embeddings. Following the previous work~\cite{wu2023billion}, we use the cross entropy (CE)~\cite{bishop2006pattern} as the loss function to maximize log probabilities for one-hop structure learning.
\begin{equation*}
    L_{\mathit{CE}} = \sum\nolimits_{i} \left( -\boldsymbol{s_{i}}\boldsymbol{e_{j}^{T}} + \mathit{log} \left( \sum\nolimits_{\mathit{e_{k} \in N\left ( s_{i} \right ) \cup e_{j}}} \boldsymbol{s_i} \boldsymbol{e_{k}^{T}} \right) \right ),
\end{equation*}
where $e_j \in s_i$ and $N(s_i) = \left \{ e_{l} \mid e_{l} \notin s_i \right \}$ denote the positive sample and the collection of negative samples, respectively.

Regarding the distillation module, the objective is to compress the dataset while persevering the knowledge as much as possible. Motivated by the previous work~\cite{zhang2024m3d}, we use the maximum mean discrepancy (MMD)~\cite{gretton2006kernel} as the loss function. The primary purpose of MMD is to determine whether two distributions are similar by comparing their samples. This is achieved by mapping the samples to a high-dimensional feature space using a kernel function and then computing the mean distance between these features. This function is particularly useful in transfer learning~\cite{long2017deep}, which needs to quantify the difference between two sets of data.
\begin{equation*}
    \resizebox{\linewidth}{!}{%
    $L_{\mathit{MMD}} = \frac{1}{n^2} \sum_{i,j} k(\boldsymbol{S_{\mathit{ob}}^{i}}, \boldsymbol{S_{\mathit{ob}}^{j}}) + \frac{1}{m^2} \sum_{i,j} k(\boldsymbol{S_{\mathit{cb}}^{i}}, \boldsymbol{S_{\mathit{cb}}^{j}}) - \frac{2}{nm} \sum_{i,j} k(\boldsymbol{S_{\mathit{ob}}^{i}}, \boldsymbol{S_{\mathit{cb}}^{j}})$,
    }
\end{equation*}
where $k$ is a kernel function (e.g., the Gaussian kernel) while $n$ and $m$ denote the size of batch data $\boldsymbol{S_{\mathit{ob}}}$ and $\boldsymbol{S_{\mathit{cb}}}$, respectively.

To train these models, we first split the underlying data based on the batch size $B_{d}$. Then, the data batches are divided into two parts, the training dataset and the testing dataset. Since the element universe is finite and fixed, we create the fixed representation with dimension $M \times d$, where each row represents one element. For each training batch, we propose a hybrid training method that minimizes an overall loss function $L$, combining $L_{CE}$ and $L_{MMD}$. To prevent overfitting, we also use the L2 regularization technique.
% , resulting in the final optimization objective.
\begin{equation*}
    \mbox{min} \left ( L + \lambda L2_{\mathit{reg}} \right ) =
    \mbox{min} \left ( L_{\mathit{CE}} + L_{\mathit{MMD}} + \lambda \left \| \Theta  \right \| ^ {2} \right ),
\end{equation*}
where $\lambda$ is the hyper-parameter to adjust the weight. 
% We also evaluate the evaluation dataset using $L$ to obtain the optimized parameters with the smallest error.

\section{Analyzer Design}
\label{sec5}
Given a query $q$ and the distilled dataset $\boldsymbol{S_c}$,  \name discovers the relations between query elements and data. Then, we capture the correlation between query elements. The key challenge is the attention mechanism~\cite{vaswani2017attention} used in \name. To handle the variable-size input, we also propose an attention-based pooling method. Finally, we employ a linear regression model to predict the cardinality of $q$, taking the fixed-size embedding as the input. Detailed explanations are provided in Sections~\ref{sec5.1} and~\ref{sec5.2}.

\subsection{Element Correlation}
\label{sec5.1}
Before estimating the cardinality of a query $q$, we need to obtain the query representation. 
A na\"ive method is to leverage the trained aggregator to integrate the information from query elements. 
Because the element embeddings are randomly initialized and fixed in the data encoder, these initial embeddings lack meaningful information.
Additionally, the aggregator cannot capture the correlation between query elements $\{e_i \mid e_i \in q \}$. Therefore, we need to propose another method to complete the task.

Considering Figure~\ref{fig:graph2}, each element can also be represented as the collection of sets containing it. Thus, we propose to learn the query representation from the underlying data. 
\textcolor{revision}{A simple method is to flatten $\boldsymbol{S_{c}}$ into a vector and concatenate the vector with embeddings of query elements. Then, the vector combining data and query information can be fed into an MLP to generate the query embedding. However, we observe that an element has a stronger relation to the sets containing it. Therefore, we leverage the cross-attention mechanism, which can pay more attention to these sets and learn better embeddings of query elements based on the distilled dataset representation.}

\textcolor{revision}{After obtaining the embeddings, we need to capture the correlations hidden in the embeddings. A simple approach is to use an MLP to learn the correlations dynamically. However, an MLP applies the same transformation to all inputs regardless of their importance and struggles to capture complex correlations.}
Motivated by the previous work~\cite{vaswani2017attention, li2023alece}, we utilize the self-attention mechanism, taking these latent embeddings as the input, to capture the correlations between elements.

\begin{figure}[htb]
    \centering
    \includegraphics[width=.9\linewidth]{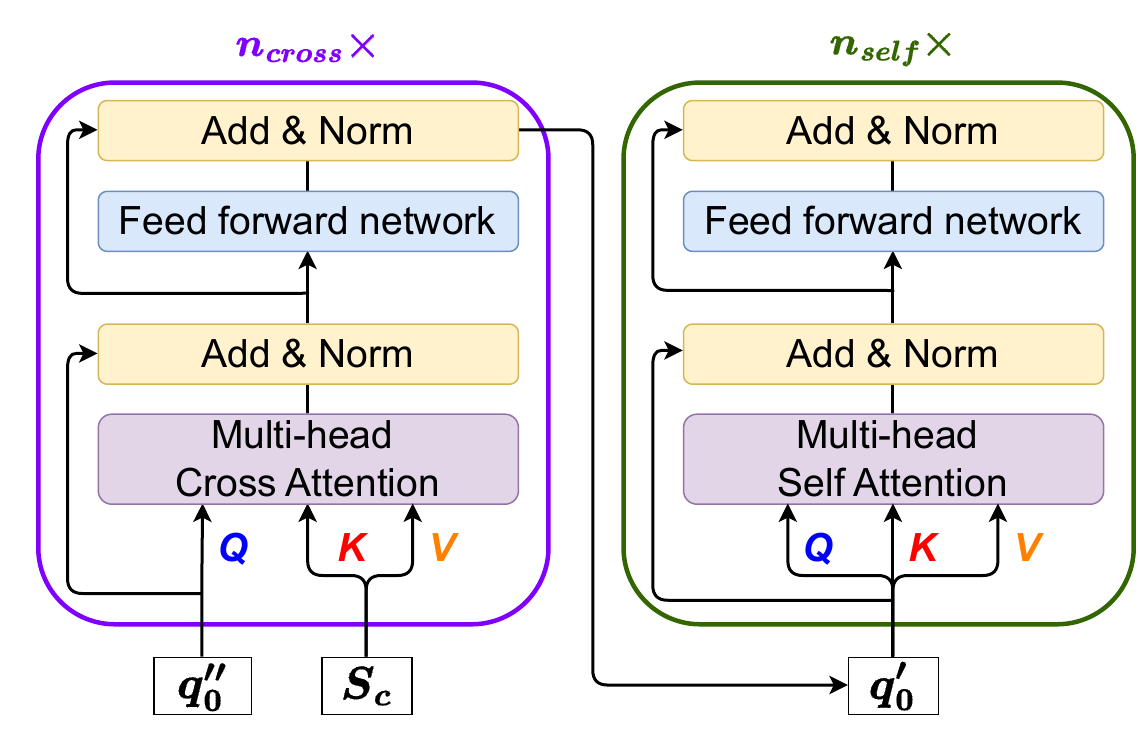}
    \vspace{-0.3cm}
    \caption{Hybrid attention framework.}
    \Description{The hybrid framework uses cross-attention and self-attention to capture the correlation between query elements}
    \label{fig:ana}
\end{figure}

In the first stage, we initialize the query embedding by stacking query element embeddings and update the query embedding by considering the information from the dataset. 
We employ $n_{cross}$ stacked attention layers to capture the correlations between the initial query embedding $\boldsymbol{q''_{0}}$ and the distilled data matrix $\boldsymbol{S_{c}}$.
%The first module takes the initial query embedding $\boldsymbol{q''_{0}}$ and the distilled data matrix $\boldsymbol{S_{c}}$ as inputs and feeds them into an $n_{cross}$-stacked layers. 
Each layer is identical and includes two sub-layers. The first is the multi-head cross-attention sub-layer $Att_{c}$ where $\boldsymbol{S_{c}}$ is used as $\boldsymbol{K}$ and $\boldsymbol{V}$ while $\boldsymbol{Q}$ uses $\boldsymbol{q''_{0}}$ or the output of the last layer. On top of $Att_{c}$, the feed-forward sub-layer $FFN$ uses stacked fully connected networks and nonlinear activation functions, e.g., GeGLU~\cite{shazeer2020glu}, to map $\boldsymbol{\tilde{q}''_{i}}$ into the latent representation $\boldsymbol{q''_{i}}$. To prevent performance degradation and ease the model training, we also employ a residual connection~\cite{he2016deep}, followed by layer normalization~\cite{ba2016layer}.
\begin{equation*}
    \begin{aligned}
        \boldsymbol{\tilde{q}''_{i}} &= \mathit{LayerNorm} \left( \boldsymbol{q''_{\mathit{i-1}}} + \mathit{Att}_{c} ( \boldsymbol{q''_{\mathit{i-1}}}, \boldsymbol{S_{c}}, \boldsymbol{S_{c}} ) \right), \\
        \boldsymbol{q''_{i}} &= \mathit{LayerNorm} \left( \boldsymbol{\tilde{q}''_{i}} + \mathit{FFN} ( \boldsymbol{\tilde{q}''_{i}} ) \right)
    \end{aligned}
\end{equation*}

The attention sub-layer establishes a bridge between query elements and data. It obtains element representations by aggregating information from the most relevant parts of data embeddings $\boldsymbol{S_{c}}$, while diminishing others. The effect of particular attention can be realized through learnable parameters of different layers.
% Suppose a set-valued query contains two elements $e_{i}$ and $e_{j}$. The attention sub-layer will pay more attention to the part of the vectors in $\boldsymbol{S_{c}}$ that are relevant to $e_{i}$ and $e_{j}$. The effect of particular attention can be realized through learnable parameters of different layers.

In the second stage, we discover and measure the correlation between query elements. We stack $n_{self}$ identical attention layers. Similar to the first stage, each layer consists of a multi-head attention sub-layer and a feed-forward sub-layer. Also, residual connections are employed, followed by layer normalization. Unlike the first stage, this self-attention sub-layer $Att_{s}$ takes the same inputs of keys, values, and queries. They are either the output of the first module, denoted as $\boldsymbol{q'_{0}}$, or the previous stacked layers.
\begin{equation*}
    \begin{aligned}
        &\boldsymbol{\tilde{q}'_{i}} = \mathit{LayerNorm} \left( \boldsymbol{q'_{\mathit{i-1}}} + \mathit{Att}_{s} ( \boldsymbol{q'_{\mathit{i-1}}}, \boldsymbol{q'_{\mathit{i-1}}}, \boldsymbol{q'_{\mathit{i-1}}} ) \right), \\
        &\boldsymbol{q'_{i}} = \mathit{LayerNorm} \left( \boldsymbol{\tilde{q}'_{i}} + \mathit{FFN} ( \boldsymbol{\tilde{q}'_{i}} ) \right)
    \end{aligned}
\end{equation*}

As introduced above, $\boldsymbol{q'_{0}}$ can be considered as new embeddings for query elements, which integrate the information from the underlying dataset. The use of the same input for the keys, values, and queries makes each element in the output set of a layer attend to all outputs in the previous layer and thus attend to all elements. More importantly, the self-attention sub-layer quantitatively 'measures' the relevance between a pair of elements, enabling the effective discovery of implicit correlations between elements. Thus, the information from the data and query is encoded into the final output embedding $\boldsymbol{q'}$ that will be processed later.
% Through self-attention sub-layers, we create links among all query elements. The correlation information helpful to the cardinality estimation problem is implicitly covered and 

\subsection{Attention Pooling}
\label{sec5.2}
Through the hybrid attention framework, the query embedding $\boldsymbol{q'}$ not only links the query with the underlying dataset but also includes the correlation information of query elements, which can be used for the cardinality estimation task. A model for set-input problems should satisfy two fundamental requirements. First, it should be permutation invariant, that is, the output of the model should not change under any permutation of the elements in the input set, which is inherently satisfied by our hybrid attention framework. Second, such a model should be able to process input sets of any size. For example, if the literal of a query is composed of $k$ elements, the dimension of the output query embedding $\boldsymbol{q'}$ will be $k \times d$. Generally, three methods address this problem -- pooling, padding, and truncation. Pooling effectively reduces input size by aggregating information from local regions, thereby reducing computational load~\cite{gholamalinezhad2020pooling}. In contrast, padding increases input size, while truncation may result in the loss of important information. Additionally, pooling operations introduce a degree of translation invariance, enhancing the model's robustness to changes in the input position~\cite{tao2022pooling}. Motivated by prior works~\cite{er2016attention, touvron2021augmenting}, we employ an attention-based pooling module to generate the fixed-sized query embedding $\boldsymbol{q}$ for predicting the corresponding cardinality.

\begin{figure}[tb]
    \centering
    \includegraphics[width=.8\linewidth]{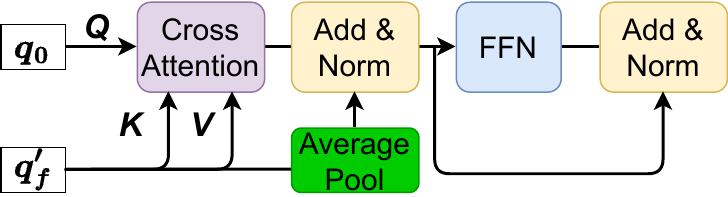}
    \vspace{-0.3cm}
    \caption{Attention pooling.}
    \Description{The attention-based pooling module for variable-size input}
    \label{fig:pool}
\end{figure}
As demonstrated in previous work~\cite{yang2019selectivity}, query elements with varying frequencies can have opposite impacts on the cardinality of a query depending on the operator type. For example, consider two queries composed of the same elements. In a superset query, which aims to find sets containing all specified elements, low-frequency elements have a stronger influence on cardinality than high-frequency elements. Conversely, in an intersection query, the resultant sets include at least one of the specified elements, meaning that high-frequency elements have a greater impact on the cardinality. Thus, the frequency information is first appended to $\boldsymbol{q'}$, which generates the $(d+1)$-dim embedding $\boldsymbol{q'_{f}}$, and then the attention pooling module takes a random initialized embedding $\boldsymbol{q_{0}}$ and $\boldsymbol{q'_{f}}$ as inputs. After accessing $\boldsymbol{q}$, the output of the pooling layer, we use a simple linear regression layer $LR$ to predicate the cardinality estimation $c$. It is noteworthy that we use the logarithm of the frequency as the appending information and modify the conventional residual connection inspired by the prior work~\cite{liu2019towards, warner2022attpool}. Figure~\ref{fig:pool} shows the framework of this module.
\begin{equation*}
    \begin{aligned}
        \boldsymbol{\tilde{q}} &= \mathit{LayerNorm} \left( \mathit{AvgPool}(\boldsymbol{q'}) + \mathit{Att}_{c} (\boldsymbol{q_{0}}, \boldsymbol{q'_{f}}, \boldsymbol{q'_{f}}) \right), \\
        \boldsymbol{q} &= \mathit{LayerNorm} \left( \boldsymbol{\tilde{q}} + \mathit{FFN}(\boldsymbol{\tilde{q}}) \right), \\
        c &= \mathit{LR}(\boldsymbol{q})
    \end{aligned}
\end{equation*}

\subsection{Analyzer Training}
\label{sec5.3}
Fine-tuning the parameters of the query analyzer requires a training dataset of which each record is a 3-tuple ($q_{i}$, $\boldsymbol{S_c}$, $c_{i}$), where $q_{i}$ is the set-valued query consisting of $k$ elements, and $c_{i}$ denotes the true cardinality of $q_{i}$. In practice, collecting the training dataset is not difficult, and we only need to collect the feedback of executed queries. The training dataset is split into batches to train our analyzer.

Since each module in the analyzer is differentiable, we train the analyzer in an end-to-end manner. Here, we use the weighted mean Q-error function $\mathit{WMQ}(\cdot)$ as the loss function, which takes input of the batch cardinality estimates $\boldsymbol{c'_{b}}$ and the true cardinalities $\boldsymbol{c_{b}}$ as well as their weights $\boldsymbol{w_{b}}$ with batch size $B_{q}$. 
\begin{equation*}
    \mathit{WMQ}(\boldsymbol{c'_{b}, \boldsymbol{c_{b}}}) = \sum\nolimits_{i=1}^{B_{q}} w_{i} * \mathit{max}\{1, \frac{c'_{i}}{c_{i}}, \frac{c_{i}}{c'_{i}}\},
\end{equation*}
where $w_{i}$ is proportional to $\log{c_{i}}$, i.e. $w_{i} = \frac{\log{c_{i}}}{\sum_j \log{c_{j}}}$. We use the weight in the loss function because it is usually beneficial to emphasize the queries with larger true cardinalities~\cite{li2023alece}.

\section{\name under Updates}
\label{sec6}
In this section, we first discuss how to leverage our \name on dynamic data. Then, we analyze its benefits compared with the state-of-the-art baseline methods. Notably, we use the same setting when working with dynamic data, that is, the element universe is finite and fixed. The update of the element universe is left for future work.

\noindent \textbf{\name on dynamic data.}
We focus on dynamic data involving insertions and deletions because one update is equivalent to one deletion followed by one insertion. Based on the structure of \name, we take a two-stage approach to accommodate dynamic data -- (1) dataset representation update and (2) query cardinality estimation.

Given a batch of tuples to be inserted, we first use the aggregator to represent them. Then, we sample the learned tuple matrix and regard sampled embeddings as the initial distilled matrix. Next, we leverage the distiller to update the distilled matrix.

When deleting tuples, considering that our original dataset is split into a collection of dataset slices based on the batch size $B_{d}$, we locate the affected slices and only need to update their corresponding distilled matrix by leveraging the trained encoder, as motivated by the previous work~\cite{bourtoule2021machine}. Additionally, we need to update the frequency of elements that are affected by the update. 

After obtaining the new distilled matrix, we can feed it along with the embeddings of the queried elements into the trained hybrid attention framework to derive the element embeddings that link the query with the updated dataset and capture the implicit correlation between elements. Subsequently, we incorporate the current frequency information of each element and utilize the attention pooling as well as the linear regression models to get the cardinality estimate for the new dataset.

\noindent \textbf{Comparison and analysis.}
When working with dynamic data, PostgreSQL reconstructs the affected histograms to approximate the distribution of the updated dataset. However, it still relies on the (partially) independent assumption, which limits its accuracy. The update process of traditional sampling methods involves sampling tuples from the inserted dataset or deleting tuples from the existing samples when encountering insertions or deletions, respectively, leading to their performance heavily depending on the quality of the resultant samples. Additionally, they still pay more attention to elements with high frequency and achieve poor performance on the query with low-frequency elements.

One prior work~\cite{yang2019selectivity} proposes the improved sampling method based on the pre-constructed trie structure. However, this study does not address how to handle dynamic data updates. Therefore, we propose a straightforward algorithm to support such updates. When deleting data, we adhere to the traditional method by checking if the data is part of the sampling results. If it is, we delete it and re-sample some sets to maintain the sample ratio. When inserting data, updating the trie structure is not feasible because it only retains the most frequent elements. Instead, we first partition data into several clusters based on the elements they contain. Then, we use the sample ratio calculated by the original trie to sample additional sets and update the sampling results. Nonetheless, this method may not well approximate the distribution of elements because of the fixed trie structure.
Another recent work~\cite{meng2023selectivity} proposes two conversion methods that transform the set-valued data into a small number of categorical data and introduces incremental updating methods for dynamic data. However, a significant issue with the proposed method persists. The cluster generation process is based on the dataset before any updates, aiming to alleviate the effect of the correlation between elements within the same cluster. As analyzed in Section~\ref{sec5}, the correlation between elements is influenced by the corresponding dataset. Thus, the clusters need to be monitored and reconstructed when necessary because the initial clusters might not work well, which the proposed methods ignore.

Compared to these baselines, the performance of our \name is superior as the data encoder minimizes the information loss when representing the dataset and the query analyzer effectively captures the useful correlation to obtain accurate estimates. 
% Moreover, our method is efficient because it only requires the feed-forward process.

\section{Experiments}
\label{sec7}
This section reports the experiments that compare \name with SOTA baselines. All experiments are evaluated on the Katana server~\cite{PVC} with a 32-core Xeon(R) Gold 6242 CPU @ 2.80GHz, 100GB memory, and an NVIDIA Tesla V100-SXM2 32GB GPU.

\subsection{Datasets and Workloads}
\label{sec7.1}
\noindent \textbf{Datasets.}
We use three real-world datasets varying in the number of sets $N$, the size of the element universe $M$, and the average number of elements within a set $\mathit{AvgL}$ as described in Table~\ref{tab:data}. The \textbf{GN} dataset~\cite{meng2023selectivity} contains descriptions of natural features, canals, and reservoirs in the United States, each of which might consist of its name, class, and location state. The \textbf{WIKI} dataset~\cite{vrandevcic2014wikidata} consists of the first sentence of each English Wikipedia article extracted in September 2017. The \textbf{TW} dataset~\cite{chen2015temporal} includes tweets posted from April 2012 to December 2012, which are published in UCR STAR \cite{GVE+19}. We preprocess the latter two datasets and convert each set to a set of words that do not include stop words.

\begin{table}[htb]
\small
\caption{Dataset statistics}
\vspace{-0.3cm}
\begin{tabular}{|c|c|c|c|}
\hline
\textbf{Property} & \textbf{GN} & \textbf{WIKI} & \textbf{TW} \\ \hline
$N$                 & 2.2M        & 5.3M        & 19.9M        \\ \hline
$M$                 & 89K         & 858K        & 559K         \\ \hline
$\mathit{AvgL}$              & 3           & 12          & 5             \\ \hline
\end{tabular}
\label{tab:data}
\end{table}

\noindent \textbf{Workloads.}
We follow the method from the former work~\cite{meng2023selectivity} to generate our workloads. For each subset query, we uniformly draw 5–10 sets from the set-valued dataset and take the union of the sampled sets as the query. For each superset and overlap query, we uniformly draw one set from the dataset, and then uniformly draw 2–4 elements from the set as the query. We also consider the frequency of query elements. Following the former work~\cite{sheng2023wisk}, we separate elements into three classes based on their frequency: low ($\leq 0.01 \%$), medium, and high ($\geq 0.1 \%$). By default, all elements are considered in a regular query. For high-frequency queries, we add a filter to select only high-frequency elements. The generation of low-frequency queries follows a similar approach. The true cardinality of each query is obtained by executing it in PostgreSQL. For each dataset, we generate \textcolor{revision}{1400} queries as the training workload, where the ratio of regular, high-frequency, and low-frequency queries is 3:2:2, while each testing workload consists of 300 queries.

\subsection{Experimental Settings}
\noindent \textbf{Implementations.}
Our \name is implemented with PyTorch~\cite{paszke2019pytorch} and we set the embedding dimension $d = 64$. We train \name using Adam optimizer~\cite{kingma2014adam}, with a learning rate of 0.001. We set the data batch size $B_d = 10000$, the distillation ratio $r = 0.001$, and the query batch size $B_q = 100$. The number of layers in the distillation model $n_{\mathit{distill}}$, the cross-attention module $n_{\mathit{cross}}$, and the self-attention module $n_{\mathit{self}}$ are set to 4, 4, and 8, respectively. When utilizing the multi-head attention mechanism, we follow the existing work~\cite{li2023alece} by setting the number of heads to 8. For all datasets, we employ negative sampling with 10 samples for each set and perform a grid search for the L2 regularization weight $\lambda \in [0, 0.005]$.

\noindent \textbf{Competitors.}
We include the following representative methods. (1)~\textbf{PG} is the 1-D histogram-based cardinality estimator used in PostgreSQL~\cite{korotkov2016selectivity}. (2)~\textbf{Sampling} uniformly samples a collection of sets, where we set the sample ratio as 0.01. (3)~\textbf{Greek-S}~\cite{moerkotte2020alpha} proposes a different method to calculate the caridnality based on the statictis of the sampled dataset. (4)~\textbf{OT-S}~\cite{yang2019selectivity} samples a collection of sets based on the constructed trie structure. For a pair comparison, the sampling ratio is the same as the previous approach and we keep the 12 most frequent elements following the setting in the previous work. (5)~\textbf{Set-Trans}~\cite{lee2019set} is proposed to capture element correlation and trained in a supervised learning manner. We regard it as a query-driven estimator. (6)~\textbf{ST} and \textbf{STH}~\cite{meng2023selectivity} convert the set-valued set into a small number of numerical data and employ the existing estimators. Based on the observation of the former work, we use DeepDB~\cite{deepdb} and NeuroCard~\cite{neurocard} as the employed estimators for \textbf{ST} and \textbf{STH}, respectively.

\noindent \textbf{Evaluation metrics.}
We use four metrics to evaluate all methods. (1) \textbf{Q-error}~\cite{moerkotte2009preventing} measures the distance between the estimated cardinality $c_{p}$ and the true cardinality $c$ of a query. In particular, $\mbox{Q-error} = max\{1, \frac{c_{p}}{c}, \frac{c}{c_{p}}\}$. (2) \textbf{Building time} denotes 
\textcolor{revision}{the construction time of traditional methods or the training time of Set-Trans and~\name. For ST and STH, the building time consists of conversion time as well as offline training time}. (3) \textbf{Storage overhead} is the memory size used by a method. (4) \textbf{Estimation latency} is the average estimation time per query.

\subsection{Overall Performance}
\label{sec7.3}
We first conduct extensive experiments to evaluate the overall performance. We do not compare ST and STH on the overlap query since they are incompatible with this query type.

\noindent \textbf{\uline{Estimation Accuracy.}}
Tables~\ref{tab:sup} -~\ref{tab:over} show the estimation error for various queries. We observe \name has the best performance compared to other baselines in most cases. The mean Q-error of \name in all cases is smaller than 10. In contrast, none of the other methods can reach this level of performance. Additionally, at the 95\% quantile, PG, Sampling, OT-S, ST and STH averagely result in up to 16.7$\times$, 29.6$\times$, 27.7$\times$, 13.5$\times$, 10.2$\times$ larger Q-error than that of \name, respectively. Next, we analyze each method individually.
\begin{table}[htb]
\caption{Estimation error for subset queries}
\vspace{-0.3cm}
\label{tab:sub}
\resizebox{\linewidth}{!}{%
\begin{tabular}{|c|c|cccc|cccc|cccc|}
\hline
\multicolumn{1}{|l|}{\multirow{2}{*}{Dataset}} & \multirow{2}{*}{Method} & \multicolumn{4}{c|}{Regular}                                  & \multicolumn{4}{c|}{High-frequency}                           & \multicolumn{4}{c|}{Low-frequency}                            \\ \cline{3-14} 
\multicolumn{1}{|l|}{}                         &                         & Mean          & 50\%          & 95\%          & 99\%          & Mean          & 50\%          & 95\%          & 99\%          & Mean          & 50\%          & 95\%          & 99\%          \\ \hline
\multirow{8}{*}{GN}                            & PG                      & 8.53          & 6.54          & 16.9          & 33.2          & 4.12          & 3.87          & 6.03          & 10.6          & \textbf{2.75}    & \textbf{2.25}    & \textbf{6}       & \textbf{9}       \\
                                               & Sampling                & 1.36          & 1.21          & 2.54          & {\ul 3.91}          & {\ul 1.11}    & {\ul 1.09}    & {\ul 1.31}    & {\ul 1.45}    & 62.3          & 13            & 166           & 203           \\
                                               & Greek-S                 & 1.39          & 1.14          & {\ul 1.75}          & 8.48          & 1.32          & 1.16          & 1.87          & 5.48          & 18.2          & 14.1          & 44.4          & 53.3          \\
                                               & OT-S                    & \textbf{1.08} & \textbf{1.06} & \textbf{1.21} & \textbf{1.32} & \textbf{1.09} & \textbf{1.07} & \textbf{1.23} & \textbf{1.32} & 63.7          & 13            & 170           & 203           \\
                                               & Set-Trans               & \textcolor{revision}{1.51}         & \textcolor{revision}{1.33}          & \textcolor{revision}{2.46}          & \textcolor{revision}{3.56}          & \textcolor{revision}{1.77}          & \textcolor{revision}{1.42}          & \textcolor{revision}{4.14}          & \textcolor{revision}{6.57}          & \textcolor{revision}{134}           & \textcolor{revision}{111}           & \textcolor{revision}{325}           & \textcolor{revision}{513}           \\
                                               & ST                      & 3.81          & 2.73          & 9.53          & 18.9          & 6.06          & 4.77          & 14.2          & 25.1          & 21.2          & 18            & 48.1          & 66.1          \\
                                               & STH                     & 2.75          & 2.46          & 5.69          & 8.67          & 1.12          & {\ul 1.09}    & 1.35          & 1.46          & 17.1          & 20.5          & 32.1          & 54.5          \\
                                               & \name                   & \textcolor{revision}{\ul 1.26}    & \textcolor{revision}{\ul 1.13}    & \textcolor{revision}{2.34}    & \textcolor{revision}{4.17}    & \textcolor{revision}{1.69}          & \textcolor{revision}{1.44}          & \textcolor{revision}{3.26}          & \textcolor{revision}{5.22}          & \textcolor{revision}{\ul 3.11} & \textcolor{revision}{\ul 2.81} & \textcolor{revision}{\ul 8.56} & \textcolor{revision}{\ul 12.4} \\ \hline
\multirow{8}{*}{WIKI}                          & PG                      & 9.49          & 5.68          & 19.7          & 41.3          & 4.74          & 3.51          & 11.9          & 14.6          & {\ul 4.14}    & {\ul 3.83}    & {\ul 7.67}    & {\ul 9.51}    \\
                                               & Sampling                & 28.1          & {\ul 1.37}    & 188           & 299           & 24.5          & 1.41          & 151           & 207           & 28.7          & 22.6          & 54.5          & 89            \\
                                               & Greek-S                 & {\ul 2.24}    & 1.85          & {\ul 5.13}    & {\ul 8.38}    & {\ul 2.73}    & {\ul 1.82}    & 7.24          & {\ul 10.7}    & 25.6          & 18.1          & 45.5          & 53.4          \\
                                               & OT-S                    & 25.3          & 1.45          & 109           & 215           & 21.7          & 1.61          & 131           & 189           & 30.7          & 23.3          & 52.1          & 84            \\
                                               & Set-Trans               & \textcolor{revision}{2.84}          & \textcolor{revision}{1.62}          & \textcolor{revision}{5.97}          & \textcolor{revision}{11.2}         & \textcolor{revision}{2.83}          & \textcolor{revision}{2.14}          & \textcolor{revision}{\ul 6.19}    & \textcolor{revision}{12.4}          & \textcolor{revision}{6.97}          & \textcolor{revision}{5.25}          & \textcolor{revision}{17.1}          & \textcolor{revision}{37.8}          \\
                                               & ST                      & 7.39          & 1.78          & 7.05          & 14.7          & 5.39          & 2.33          & 28.9          & 35.3          & 13.8          & 7.11          & 54            & 135           \\
                                               & STH                     & 7.32          & 5.32          & 19.7          & 30.8          & 10.1          & 8.23          & 22.8          & 33.9          & 12.4          & 10.2          & 48.6          & 89.2          \\
                                               & \name                   & \textcolor{revision}{\textbf{2.04}} & \textcolor{revision}{\textbf{1.36}} & \textcolor{revision}{\textbf{4.93}} & \textcolor{revision}{\textbf{8.71}} & \textcolor{revision}{\textbf{2.37}} & \textcolor{revision}{\textbf{1.77}} & \textcolor{revision}{\textbf{5.35}} & \textcolor{revision}{\textbf{7.27}} & \textcolor{revision}{\textbf{2.43}} & \textcolor{revision}{\textbf{1.75}} & \textcolor{revision}{\textbf{5.96}} & \textcolor{revision}{\textbf{13.5}} \\ \hline
\multirow{8}{*}{TW}                            & PG                      & 5.85          & 4.39          & 13.8          & 23.2          & 4.01          & 3.19          & 7.74          & 12.2          & {\ul 3.69}    & {\ul 2.88}    & {\ul 9.13}    & {\ul 12.7}    \\
                                               & Sampling                & 5.04          & 4.03          & 12.1          & 36.2          & 3.57          & 3.24          & 13.5          & 29.1          & 19.1          & 15.4          & 52            & 76            \\
                                               & Greek-S                 & {\ul 1.81}          & {\ul 1.42}    & {\ul 2.49}    & {\ul 3.92}          & {\ul 1.93}    & {\ul 1.81}    & {\ul 3.54}    & 11.9          & 14.4          & 11.5          & 38.1          & 53.5          \\
                                               & OT-S                    & 4.99          & 3.83          & 9.67          & 28.4          & 4.82          & 3.55          & 11.6          & 31.2          & 21.9          & 16.2          & 60            & 97            \\
                                               & Set-Trans               & \textcolor{revision}{2.05}    & \textcolor{revision}{1.89}          & \textcolor{revision}{3.72}          & \textcolor{revision}{4.47}    & \textcolor{revision}{2.74}          & \textcolor{revision}{2.54}          & \textcolor{revision}{5.81}          & \textcolor{revision}{\ul 9.58}    & \textcolor{revision}{375}           & \textcolor{revision}{260}           & \textcolor{revision}{922}           & \textcolor{revision}{1464}          \\
                                               & ST                      & 4.38          & 3.74          & 8.82          & 12.4          & 4.74          & 3.31          & 7.32          & 15.4          & 30.9          & 23.1          & 74            & 127           \\
                                               & STH                     & 5.07          & 3.24          & 12.3          & 15.2          & 3.11          & 2.76          & 5.94          & 11.1          & 19.5          & 4.57          & 72.7          & 130           \\
                                               & \name                   & \textcolor{revision}{\textbf{1.46}} & \textcolor{revision}{\textbf{1.34}} & \textcolor{revision}{\textbf{2.17}} & \textcolor{revision}{\textbf{2.81}} & \textcolor{revision}{\textbf{1.66}} & \textcolor{revision}{\textbf{1.53}} & \textcolor{revision}{\textbf{2.94}} & \textcolor{revision}{\textbf{4.24}} & \textcolor{revision}{\textbf{1.88}} & \textcolor{revision}{\textbf{1.61}} & \textcolor{revision}{\textbf{3.81}} & \textcolor{revision}{\textbf{4.92}} \\ \hline
\end{tabular}%
}
\end{table}

\begin{table}[htb]
\caption{Estimation error for superset queries}
\vspace{-0.3cm}
\label{tab:sup}
\resizebox{\linewidth}{!}{%
\begin{tabular}{|c|c|cccc|cccc|cccc|}
\hline
\multicolumn{1}{|l|}{\multirow{2}{*}{Dataset}} & \multirow{2}{*}{Method} & \multicolumn{4}{c|}{Regular}                                  & \multicolumn{4}{c|}{High-frequency}                           & \multicolumn{4}{c|}{Low-frequency}                            \\ \cline{3-14} 
\multicolumn{1}{|l|}{}                         &                         & Mean          & 50\%          & 95\%          & 99\%          & Mean          & 50\%          & 95\%          & 99\%          & Mean          & 50\%          & 95\%          & 99\%          \\ \hline
\multirow{8}{*}{GN}                            & PG                      & 67.5          & 4.26          & 198           & 1785          & 96.6          & 3.5           & 192           & 2728          & 6.73          & 6             & 12            & 17            \\
                                               & Sampling                & 16.7          & 5             & 70            & 187           & 21.1          & 2.45          & 94            & 229           & 37.1          & 35.2          & 76.7          & 90.2          \\
                                               & Greek-S                 & {\ul 12.3}          & 5.93          & 44.4          & {\ul 53.3}    & {\ul 8.05}          & 3.46          & 33.3          & {\ul 53.3}    & 25.8          & 17.4          & 50.6          & 53.3          \\
                                               & OT-S                    & 20.2          & 6             & 81            & 195           & 17.9          & 2.21          & 78            & 192           & 36.7          & 34.5          & 77.1          & 91.3          \\
                                               & Set-Trans               & \textcolor{revision}{12.5}    & \textcolor{revision}{5.77}          & \textcolor{revision}{46.9}          & \textcolor{revision}{117}           & \textcolor{revision}{12.4}          & \textcolor{revision}{4.32}          & \textcolor{revision}{46.5}          & \textcolor{revision}{133}           & \textcolor{revision}{7.53}    & \textcolor{revision}{5.22}    & \textcolor{revision}{\ul 18.6}    & \textcolor{revision}{33.5}          \\
                                               & ST                      & 40.2          & 2.21          & 52.7          & 272           & 18.8          & {\ul 2.12}          & 22.2          & 100           & 9.71          & 7             & 28.5          & 37.4          \\
                                               & STH                     & 16.3          & \textbf{1.52} & {\ul 34.6}    & 161           & 10.3          & \textbf{1.45} & {\ul 22.2}    & 150           & {\ul 5.08}          & {\ul 5}             & 11            & {\ul 15}      \\
                                               & \name                   & \textcolor{revision}{\textbf{5.19}} & \textcolor{revision}{2.91}    & \textcolor{revision}{\textbf{17.2}} & \textcolor{revision}{\textbf{36.1}} & \textcolor{revision}{\textbf{6.54}} & \textcolor{revision}{2.19}    & \textcolor{revision}{\textbf{20.9}} & \textcolor{revision}{\textbf{49.6}} & \textcolor{revision}{\textbf{2.15}} & \textcolor{revision}{\textbf{2.11}} & \textcolor{revision}{\textbf{3.18}} & \textcolor{revision}{\textbf{3.49}} \\ \hline
\multirow{8}{*}{WIKI}                          & PG                      & 175           & 8             & 271           & 2330          & 439           & 8.61          & 703           & 5479          & {\ul 9.44}    & {\ul 3.75}    & 33            & 89            \\
                                               & Sampling                & 13.3          & \textbf{2.31}    & 65            & 145           & 15.1          & \textbf{1.39} & 71            & 179           & 32.9          & 28.1          & 83            & 189           \\
                                               & Greek-S                 & {\ul 10.9}    & 3.98          & {\ul 38.1}    & {\ul 53.4}    & {\ul 8.91}    & 2.35          & {\ul 33.4}    & {\ul 93.5}    & 15.4          & 12.4          & {\ul 30.4}    & {\ul 53.5}    \\
                                               & OT-S                    & 14.8          & 2.57          & 72            & 159           & 11.6          & {\ul 1.51}    & 61            & 159           & 33.7          & 29.3          & 77            & 186           \\
                                               & Set-Trans               & \textcolor{revision}{19.2}          & \textcolor{revision}{7.86}          & \textcolor{revision}{80.2}          & \textcolor{revision}{158}           & \textcolor{revision}{15.1}          & \textcolor{revision}{5.43}          & \textcolor{revision}{60.2}          & \textcolor{revision}{108}           & \textcolor{revision}{22.6}          & \textcolor{revision}{10.7}          & \textcolor{revision}{81.8}          & \textcolor{revision}{96.1}          \\
                                               & ST                      & 130           & 4.92          & 245           & 1570          & 130           & 2.77          & 101           & 1576          & 30.6          & 13.3          & 118           & 241           \\
                                               & STH                     & 16.1          & 3.98          & 69.6          & 169           & 9.31          & 2.58          & 35.4          & 143           & 29.3          & 11            & 118           & 239           \\
                                               & \name                   & \textcolor{revision}{\textbf{6.44}} & \textcolor{revision}{\ul 3.34} & \textcolor{revision}{\textbf{17.9}} & \textcolor{revision}{\textbf{37.4}} & \textcolor{revision}{\textbf{8.33}} & \textcolor{revision}{3.16}         & \textcolor{revision}{\textbf{30.8}} & \textcolor{revision}{\textbf{75.2}} & \textcolor{revision}{\textbf{2.88}} & \textcolor{revision}{\textbf{2.67}} & \textcolor{revision}{\textbf{8.23}} & \textcolor{revision}{\textbf{11.6}} \\ \hline
\multirow{8}{*}{TW}                            & PG                      & 179           & 2.39          & 99            & 1014          & 128           & 2.09          & 48            & 2323          & 14.3          & {\ul 4.75}    & 53.5          & 152           \\
                                               & Sampling                & 17.5          & 2.31          & 83            & 223           & 10.7          & \textbf{1.32} & 60            & 173           & 173           & 149           & 251           & 380           \\
                                               & Greek-S                 & {\ul 9.83}    & 3.21          & {\ul 44.5}    & {\ul 74.3}    & {\ul 9.22}    & 2.43          & {\ul 26.7}    & {\ul 80.3}    & 22.7          & 19.1          & 47.6          & {\ul 53.3}    \\
                                               & OT-S                    & 15.6          & \textbf{1.96} & 87            & 210           & 8.91          & {\ul 1.36}    & 44            & 135           & 161           & 137           & 251           & 380           \\
                                               & Set-Trans               & \textcolor{revision}{20.2}          & \textcolor{revision}{6.67}          & \textcolor{revision}{82.7}          & \textcolor{revision}{128}           & \textcolor{revision}{13.3}          & \textcolor{revision}{4.67}          & \textcolor{revision}{65.6}          & \textcolor{revision}{118}           & \textcolor{revision}{\ul 13.9}    & \textcolor{revision}{10.5}          & \textcolor{revision}{\ul 37.2}    & \textcolor{revision}{60.7}          \\
                                               & ST                      & 49.9          & 2.37          & 81.1          & 578           & 43.7          & 2.05          & 76.6          & 553           & 37.9          & 10            & 160           & 347           \\
                                               & STH                     & 12.2          & {\ul 2.08}    & 48.6          & 306           & 11.7          & 2.83          & 44.1          & 471           & 37.7          & 10.1          & 160           & 374           \\
                                               & \name                   & \textcolor{revision}{\textbf{6.79}} & \textcolor{revision}{2.32}          & \textcolor{revision}{\textbf{22.5}} & \textcolor{revision}{\textbf{60.7}} & \textcolor{revision}{\textbf{8.41}} & \textcolor{revision}{2.04}          & \textcolor{revision}{\textbf{26.8}} & \textcolor{revision}{\textbf{67.1}} & \textcolor{revision}{\textbf{3.93}} & \textcolor{revision}{\textbf{2.61}} & \textcolor{revision}{\textbf{8.76}} & \textcolor{revision}{\textbf{11.7}} \\ \hline
\end{tabular}%
}
\end{table}

\begin{table}[htb]
\caption{Estimation error for overlap queries}
\vspace{-0.3cm}
\label{tab:over}
\resizebox{\linewidth}{!}{%
\begin{tabular}{|c|c|cccc|cccc|cccc|}
\hline
\multicolumn{1}{|l|}{\multirow{2}{*}{Dataset}} & \multirow{2}{*}{Method} & \multicolumn{4}{c|}{Regular}                                  & \multicolumn{4}{c|}{High-frequency}                           & \multicolumn{4}{c|}{Low-frequency}                            \\ \cline{3-14} 
\multicolumn{1}{|l|}{}                         &                         & Mean          & 50\%          & 95\%          & 99\%          & Mean          & 50\%          & 95\%          & 99\%          & Mean          & 50\%          & 95\%          & 99\%          \\ \hline
\multirow{6}{*}{GN}                            & PG                      & \textbf{3.28}    & \textbf{1.26}    & {\ul 9.82}    & 34.1          & 2.99          & \textbf{1.31}    & 9.47          & {\ul 27.4}    & {\ul 12.7}    & 4.15          & 68.1          & {\ul 108}     \\
                                               & Sampling                & 4.18          & 1.28          & 9.95          & 42.1          & {\ul 2.93}    & {\ul 1.32}          & 9.19          & 29.2          & 239           & 2.89          & 481           & 5954          \\
                                               & Greek-S                 & 4.99          & 1.33          & 11.1          & 75.2          & 2.99          & 1.35          & {\ul 8.66}    & 28.1          & 13.6          & {\ul 2.71}    & {\ul 33.3}    & 277           \\
                                               & OT-S                    & 3.41          & 1.28          & 9.96          & {\ul 30.6}    & 3.01          & 1.33          & 8.81          & 29.8          & 235           & 3.08          & 502           & 6409          \\
                                               & Set-Trans               & \textcolor{revision}{10.7}          & \textcolor{revision}{4.73}          & \textcolor{revision}{46.2}          & \textcolor{revision}{110}           & \textcolor{revision}{10.9}          & \textcolor{revision}{5.51}          & \textcolor{revision}{22.1}          & \textcolor{revision}{94.7}          & \textcolor{revision}{51.2}          & \textcolor{revision}{30.3}          & \textcolor{revision}{221}           & \textcolor{revision}{346}           \\
                                               & \name                   & \textcolor{revision}{\ul 3.34} & \textcolor{revision}{1.46} & \textcolor{revision}{\textbf{9.29}} & \textcolor{revision}{\textbf{18.1}} & \textcolor{revision}{\textbf{2.66}} & \textcolor{revision}{1.56} & \textcolor{revision}{\textbf{6.39}} & \textcolor{revision}{\textbf{17.6}} & \textcolor{revision}{\textbf{9.62}} & \textcolor{revision}{\textbf{2.67}} & \textcolor{revision}{\textbf{22.7}} & \textcolor{revision}{\textbf{80.6}} \\ \hline
\multirow{6}{*}{WIKI}                          & PG                      & 8.28          & 1.81          & {\ul 13.1}    & 94.2          & 4.28          & 2.32          & 15.1          & 40.7          & 27.1          & 14.9          & 165           & 368           \\
                                               & Sampling                & {\ul 4.29}    & 1.88          & 13.2          & 43            & 4.28          & 2.34          & 14.2          & 37.4          & 43.5          & {\ul 2.49}    & 214           & 496           \\
                                               & Greek-S                 & 4.52          & 1.93          & 16.6          & {\ul 38.8}    & 4.31          & 2.32          & 13.5          & 38.5          & {\ul 8.44}    & 3.29          & {\ul 22.9}    & {\ul 54.8}    \\
                                               & OT-S                    & 5.52          & {\ul 1.78}    & {\ul 13.1}    & 59.3          & {\ul 4.16}    & {\ul 2.26}    & {\ul 12.9}    & {\ul 37.1}    & 35.5          & 2.54          & 196           & 414           \\
                                               & Set-Trans               & \textcolor{revision}{23.9}          & \textcolor{revision}{11.9}          & \textcolor{revision}{98.7}          & \textcolor{revision}{131}          & \textcolor{revision}{37.7}          & \textcolor{revision}{19.5}          & \textcolor{revision}{108}          & \textcolor{revision}{260}           & \textcolor{revision}{26.6}          & \textcolor{revision}{14.3}          & \textcolor{revision}{104}           & \textcolor{revision}{239}           \\
                                               & \name                   & \textcolor{revision}{\textbf{3.98}} & \textcolor{revision}{\textbf{1.75}} & \textcolor{revision}{\textbf{8.99}} & \textcolor{revision}{\textbf{15.8}} & \textcolor{revision}{\textbf{2.72}} & \textcolor{revision}{\textbf{2.14}} & \textcolor{revision}{\textbf{7.52}} & \textcolor{revision}{\textbf{10.7}} & \textcolor{revision}{\textbf{8.01}} & \textcolor{revision}{\textbf{2.46}} & \textcolor{revision}{\textbf{22.6}} & \textcolor{revision}{\textbf{38.5}} \\ \hline
\multirow{6}{*}{TW}                            & PG                      & 6.33          & 2.26          & 19.5          & 85.9          & 4.64          & 2.38          & 11.3          & 37.4          & 18.3          & 14.3          & 134           & 377           \\
                                               & Sampling                & {\ul 6.21}    & 2.26          & 19.6          & 82.1          & 4.67          & 2.42          & 10.8          & 37.6          & 28.1          & 2.83          & 175           & 473           \\
                                               & Greek-S                 & 6.37          & 2.33          & 22.4          & {\ul 76.4}    & {\ul 4.61}    & 2.41          & 11.1          & {\ul 36.2}    & {\ul 11.6}    & 3.59          & {\ul 20.6}    & 320     \\
                                               & OT-S                    & 6.25          & {\ul 2.21}    & {\ul 19.3}    & 84.1          & 4.66          & {\ul 2.35}    & {\ul 10.6}    & 37.6          & 28.6          & {\ul 2.81}    & 167           & 287           \\
                                               & Set-Trans               & \textcolor{revision}{51.1}          & \textcolor{revision}{16.5}          & \textcolor{revision}{162}          & \textcolor{revision}{446}           & \textcolor{revision}{69.2}          & \textcolor{revision}{28.9}          & \textcolor{revision}{191}          & \textcolor{revision}{580}           & \textcolor{revision}{16.2}          & \textcolor{revision}{8.68}          & \textcolor{revision}{57.7}           & \textcolor{revision}{\ul 166}           \\
                                               & \name                   & \textcolor{revision}{\textbf{5.61}} & \textcolor{revision}{\textbf{2.19}} & \textcolor{revision}{\textbf{16.4}} & \textcolor{revision}{\textbf{51.5}} & \textcolor{revision}{\textbf{2.96}} & \textcolor{revision}{\textbf{2.21}} & \textcolor{revision}{\textbf{7.92}} & \textcolor{revision}{\textbf{13.7}} & \textcolor{revision}{\textbf{6.72}} & \textcolor{revision}{\textbf{2.79}} & \textcolor{revision}{\textbf{17.9}} & \textcolor{revision}{\textbf{73.3}} \\ \hline
\end{tabular}%
}
\end{table}

PG estimates the cardinality of a query based on the independence assumption,  which often leads to poor performance on regular and high-frequency queries due to ignoring the correlation between elements. However, its estimates are more accurate for low-frequency queries, as the correlation between low-frequency elements can sometimes be disregarded. Moreover, we observe that its performance on low-frequency overlap queries is bad because of insufficient statistics targets. 
% Increasing the value could improve the accuracy but lead to higher cost.

Sampling, Greek-S and OT-S show the opposite trend compared to PG. Their performance on regular and high-frequency queries is better than that on low-frequency queries because they focus more on high-frequency elements. Greek-S firstly determines the geometric mean of the upper ($\omega$) and lower ($\alpha$) bounds for the number of qualifying tuples based on probabilistic estimates, which is then used to return a more accurate cardinality estimate. OT-S improves upon the traditional sampling method by leveraging the trie structure, leading to better performance in most cases. However, the performance of these methods is not stable across three datasets, as it heavily depends on the quality of sampling results.

Set-Trans only utilizes the information of the workload, regarding the problem as a supervised learning task. However, its performance is unstable across datasets and queries since it is impossible to enumerate all combinations given limited training data.
ST and STH are SOTA methods that can utilize any data-driven estimator to predict cardinality based on the constructed clusters and the corresponding conversion algorithm. In general, our \name outperforms these methods, verifying that the partial independence assumption in these methods is not reasonable for some scenarios. Additionally, we observe that the results on low-frequency queries differ significantly from those reported in the former work~\cite{meng2023selectivity}, as it first filters out low cardinality elements before selecting the query element, thereby ignoring the actual low-frequency elements.

\noindent \textbf{\uline{Construction Efficiency.}}
Referring to Table~\ref{tab:build}, the training time of \name is acceptable and shorter than STH and ST in most cases. It requires less than \textcolor{revision}{2, 7, and 10} minutes to fine-tune its parameters for the GN, WIKI, and TW datasets, respectively. Although Sampling and OT-S require less construction time, their Q-error performance is worse, especially on low-frequency queries. Notably, Greek-S is excluded from our analysis, as it does not influence the sampling process. Besides, we observe that the time of Set-Trans and \name on different types varies because of the variable-size query, where a subset query usually has more elements than other types of queries.

\begin{table}[htb]
\caption{Building time (minutes) of different methods}
\vspace{-0.3cm}
\label{tab:build}
\resizebox{\columnwidth}{!}{%
\begin{tabular}{|c|c|c|c|c|c|c|c|}
\hline
Dataset                 & Type     & Sampling & OT-S & Set-Trans & ST   & STH  & \name \\ \hline
\multirow{3}{*}{GN}     & Subset   & 0.03     & 0.11 & \textcolor{revision}{2.76}      & 0.43 & 4.02 & \textcolor{revision}{1.88}  \\
                        & Superset & 0.03     & 0.11 & \textcolor{revision}{2.06}      & 0.43 & 3.94 & \textcolor{revision}{1.56}  \\
                        & Overlap  & 0.03     & 0.11 & \textcolor{revision}{2.03}      & -    & -    & \textcolor{revision}{1.58}  \\ \hline
\multirow{3}{*}{WIKI}   & Subset   & 0.06     & 0.61 & \textcolor{revision}{6.03}      & 10.9 & 13.2 & \textcolor{revision}{6.77}  \\
                        & Superset & 0.06     & 0.61 & \textcolor{revision}{1.97}      & 10.9 & 12.9 & \textcolor{revision}{2.85}  \\
                        & Overlap  & 0.06     & 0.61 & \textcolor{revision}{1.68}      & -    & -    & \textcolor{revision}{2.67}  \\ \hline
\multirow{3}{*}{TW}     & Subset   & 0.21     & 1.31 & \textcolor{revision}{6.16}      & 9.73 & 11.7 & \textcolor{revision}{9.35}  \\
                        & Superset & 0.21     & 1.31 & \textcolor{revision}{5.07}      & 9.73 & 11.4 & \textcolor{revision}{5.13}  \\
                        & Overlap  & 0.21     & 1.31 & \textcolor{revision}{5.14}      & -    & -    & \textcolor{revision}{5.16}  \\ \hline
\end{tabular}%
}
\end{table}

\noindent \textbf{\uline{Storage Overhead.}}
Table~\ref{tab:size} shows the experimental results. We denote the size of samples as the storage overhead of sampling-based methods, which increases with the size of datasets. As Greek-S does not change the sampling process, its results are excluded from the table. ST and STH need to maintain the parameters of DeepDB and NeuroCard, respectively, with STH typically incurring higher space costs due to its more complex structure. In contrast, Set-Trans maintains a consistent model size across datasets, as its storage requirements are determined by the embedding dimensions. However, its overall size is slightly larger than ours due to the additional inclusion of inducing points. Our \name demonstrates a stable storage, exhibiting only a marginal increase as the dataset size expands, primarily due to the benefits of its distillation process.

\begin{table}[htb]
\caption{Storage overhead (MB) of different methods}
\vspace{-0.3cm}
\label{tab:size}
\begin{tabular}{|c|c|c|c|c|c|c|}
\hline
Dataset & Sampling & OT-S & Set-Trans & ST   & STH  & \name \\ \hline
GN      & 0.28     & 0.29 & 10.6      & 3.31 & 16.1 & 8.11  \\ \hline
WIKI    & 3.21     & 3.33 & 10.6      & 29.6 & 79.7 & 8.26  \\ \hline
TW      & 4.77     & 4.79 & 10.6      & 11.2 & 58.1 & 8.36  \\ \hline
\end{tabular}%
\end{table}

\noindent \textbf{\uline{Estimation Latency.}}
As illustrated in Table~\ref{tab:latency}, \textcolor{revision}{PG needs the least time to estimate the cardinality but its performance is not acceptable,} while the latency of the sampling-based methods increases with the size of the dataset because they need to traverse all samples. Greek-S, in particular, exhibits significantly higher latency than the other methods, as it involves more computational steps, resulting in increased time costs. As shown in previous work~\cite{meng2023selectivity}, the time complexity of ST depends on the number of clusters, which leads to lower estimation time, while STH reduces the number of nodes kept on each trie to speed up the prediction process. However, both methods need to convert the query before estimating the cardinality, which cannot be executed in GPUs. Set-Trans uses the least time to predicate the cardinality because it only takes the query as the input and regards the problem as a supervised learning task. \name is a fully learning-based estimator with the best performance and the most stable latency across three datasets.

\begin{table}[htb]
\caption{Estimation latency (ms) of different methods}
\vspace{-0.3cm}
\label{tab:latency}
\resizebox{\columnwidth}{!}{%
\begin{tabular}{|c|c|c|c|c|c|c|c|c|}
\hline
Dataset & PG    & Sampling & Greek-S & OT-S  & Set-Trans & ST    & STH   & \name \\ \hline
GN      & \textcolor{revision}{1.05} & 124.9    & 207.2   & 128.6 & 2.91      & 3.86  & 12.39 & 4.54  \\ \hline
WIKI    & \textcolor{revision}{3.64} & 381.1    & 876.1   & 392.4 & 3.28      & 25.67 & 83.12 & 5.17  \\ \hline
TW      & \textcolor{revision}{2.79}  & 1163     & 7808    & 1175  & 3.07      & 19.57 & 41.07 & 6.17  \\ \hline
\end{tabular}%
}
\end{table}

\noindent \textbf{\uline{Real-world Cases.}}
\textcolor{revision}{In Section~\ref{sec1}, we introduce a real-world application for set-valued data, i.e., tag search. Privacy policies render user data confidential in most applications, such as Twitter and Wiki. We utilize the data released by a recipe website (\url{http://www.food.com}) and its real user search queries~\cite{majumder2019generating, li2021share} to show the necessity of supporting set queries. In this dataset, the total number of keywords is 632 and there are 500K recipes, each tagged with several keywords. For the query workload, we extract 10K distinctive queries for superset and overlap queries supported by this website. For each query type, we randomly select 1000 and 200 queries for the training and validation, respectively, while the remaining are used for testing.}

\textcolor{revision}{As ST and STH cannot support the overlap queries, the two methods have no corresponding results. As shown in Table~\ref{tab:food}, \name consistently outperforms other baseline methods using the real-world set-valued data and query workload.}

\begin{table}[htb]
\caption{Real-world tag search}
\vspace{-0.3cm}
\label{tab:food}
\resizebox{\columnwidth}{!}{%
\begin{tabular}{|c|cccc|cccc|}
\hline
\multirow{2}{*}{Method} & \multicolumn{4}{c|}{Superset}                                 & \multicolumn{4}{c|}{Overlap}                                  \\ \cline{2-9} 
                        & Mean          & 50\%          & 95\%          & 99\%          & Mean          & 50\%          & 95\%          & 99\%          \\ \hline
PG                      & 15.1          & 4.25          & 51            & 186           & {\ul 1.17}    & {\ul 1.13}    & 1.74          & 2.16          \\
Sampling                & 14.4          & 4.62          & 59            & 122           & 1.26          & 1.31          & 1.88          & 3.71          \\
Greek-S                 & 4.42          & 2.31          & 17.9          & 17.9          & 1.21          & {\ul 1.13}    & 1.57          & 2.11          \\
OT-S                    & 3.68          & 2.27          & 12.5          & {\ul 12.8}    & 1.21          & 1.16          & {\ul 1.44}    & {\ul 1.98}    \\
Set-Trans               & {\ul 3.41}    & {\ul 2.19}    & {\ul 8.47}    & 20.9          & 1.61          & 1.45          & 2.24          & 4.79          \\
ST                      & 4.05          & 2.29          & 11.8          & 23.4          & -             & -             & -             & -             \\
STH                     & 10.2          & 7.07          & 29.1          & 50.6          & -             & -             & -             & -             \\
\name                   & \textbf{3.18} & \textbf{2.01} & \textbf{7.53} & \textbf{15.2} & \textbf{1.02} & \textbf{1.01} & \textbf{1.05} & \textbf{1.11} \\ \hline
\end{tabular}%
}
\end{table}

\subsection{Performance on Dynamic Data}
We follow previous studies~\cite{li2023alece, meng2023selectivity} to conduct experiments on dynamic data. We use about 70\% of the sets as the initial dataset to train our data encoder and the remaining as the collection of insertion data. The size of each insertion is equal to the data batch size $B_{d}$. 90\% of the insertion is used to train the query analyzer while the remaining is used to evaluate the performance. Additionally, we might randomly delete some sets from the current dataset before any insertions. To simulate the real-world scenarios, the number of deleted sets is only a small fraction of the entire dataset, meaning that the number of affected slices is much lower than the others. 
% As shown in the previous study~\cite{han2021cardinality}, pure query-driven approaches are not suitable for handling dynamic data scenarios. Consequently, we have excluded Set-Trans from our experimental evaluation.

Regarding the workload, we only conduct the experiments on the superset and subset query since ST and STH are incompatible with the overlap query. To train the query analyzer, we utilize the generated workload as the base workload. Then, we randomly select 20 and 10 queries from the base workload as the training and validation sets, respectively, once an insertion completes. When evaluating the performance, we generate 100 queries after any insertion as the evaluation queries of the current dataset and finally report the average value. Note that the true cardinality of a query might change due to the dynamic data. Thus, we need to use PostgreSQL to obtain the true cardinality values and filter the queries without any results in the training or evaluation process.

\begin{table}[htb]
\caption{The performance on dynamic data}
\vspace{-0.3cm}
\label{tab:dynamic}
\resizebox{\columnwidth}{!}{%
\begin{tabular}{|c|c|ccccc|ccccc|}
\hline
\multirow{3}{*}{Dataset} & \multirow{3}{*}{Method} & \multicolumn{5}{c|}{Superset}                                                                                                                                   & \multicolumn{5}{c|}{Subset}                                                                                                                                     \\ \cline{3-12} 
                         &                         & \multicolumn{4}{c|}{Q-error}                                                       & \multirow{2}{*}{\begin{tabular}[c]{@{}c@{}}Update\\ Time (s)\end{tabular}} & \multicolumn{4}{c|}{Q-error}                                                       & \multirow{2}{*}{\begin{tabular}[c]{@{}c@{}}Update\\ Time (s)\end{tabular}} \\ \cline{3-6} \cline{8-11}
                         &                         & Mean          & 50\%          & 95\%          & \multicolumn{1}{c|}{99\%}          &                                                                            & Mean          & 50\%          & 95\%          & \multicolumn{1}{c|}{99\%}          &                                                                            \\ \hline
\multirow{8}{*}{GN}      & PG                      & 69.6          & 5.01          & 204           & \multicolumn{1}{c|}{2031}          & -                                                                          & 8.39          & 7.23          & 15.1          & \multicolumn{1}{c|}{35.7}          & -                                                                          \\
                         & Sampling                & 18.7          & 7             & 84            & \multicolumn{1}{c|}{173}           & 0.01                                                                       & \textbf{1.37} & \textbf{1.16} & 3.18          & \multicolumn{1}{c|}{{\ul 5.28}}    & 0.01                                                                       \\
                         & Greek-S                 & {\ul 15.4}    & 9.21          & 53.3          & \multicolumn{1}{c|}{{\ul 64.7}}    & 0.01                                                                       & 1.61          & 1.52          & {\ul 2.89}    & \multicolumn{1}{c|}{6.06}          & 0.01                                                                       \\
                         & OT-S                    & 24.2          & 7.2           & 97.2          & \multicolumn{1}{c|}{258}           & 0.06                                                                       & 1.62          & 1.59          & 2.82          & \multicolumn{1}{c|}{6.98}          & 0.06                                                                       \\
                         & Set-Trans                   & \textcolor{revision}{28.8}          & \textcolor{revision}{11.3}           & \textcolor{revision}{78.2}          & \multicolumn{1}{c|}{\textcolor{revision}{211}}           & \textcolor{revision}{0.26}                                                                       & \textcolor{revision}{5.77}          & \textcolor{revision}{4.83}           & \textcolor{revision}{7.18}          & \multicolumn{1}{c|}{\textcolor{revision}{9.49}}           & \textcolor{revision}{0.34}                                                                       \\
                         & ST                      & 45.7          & 4.35          & 78.1          & \multicolumn{1}{c|}{395}           & 0.16                                                                       & 4.51          & 3.17          & 12.7          & \multicolumn{1}{c|}{20.2}          & 0.16                                                                       \\
                         & STH                     & 20.5          & {\ul 4.02}    & {\ul 49.9}    & \multicolumn{1}{c|}{190}           & 0.17                                                                       & 3.88          & 2.56          & 8.76          & \multicolumn{1}{c|}{11.5}          & 0.17                                                                       \\
                         & \name                   & \textbf{5.35} & \textbf{3.09} & \textbf{15.3} & \multicolumn{1}{c|}{\textbf{42.7}} & 0.34                                                                       & {\ul 1.59}    & {\ul 1.51}    & \textbf{2.77} & \multicolumn{1}{c|}{\textbf{5.01}} & 0.41                                                                       \\ \hline
\multirow{8}{*}{WIKI}    & PG                      & 136           & 6.68          & 341           & \multicolumn{1}{c|}{3249}          & -                                                                          & 10.1          & 4.93          & 18.2          & \multicolumn{1}{c|}{39.5}          & -                                                                          \\
                         & Sampling                & 14.8          & 4.01          & 71            & \multicolumn{1}{c|}{140}           & 0.01                                                                       & 29.8          & 1.95          & 197           & \multicolumn{1}{c|}{371}           & 0.01                                                                       \\
                         & Greek-S                 & {\ul 11.5}    & 4.17          & {\ul 44.5}    & \multicolumn{1}{c|}{{\ul 53.4}}    & 0.01                                                                       & {\ul 4.31}    & {\ul 1.75}    & {\ul 9.42}    & \multicolumn{1}{c|}{{\ul 13.6}}    & 0.01                                                                       \\
                         & OT-S                    & 18.5          & {\ul 3.22}    & 90            & \multicolumn{1}{c|}{199}           & 0.06                                                                       & 37.9          & 2.18          & 163           & \multicolumn{1}{c|}{323}           & 0.06                                                                       \\
                         & Set-Trans                   & \textcolor{revision}{28.1}          & \textcolor{revision}{13.7}           & \textcolor{revision}{114}          & \multicolumn{1}{c|}{\textcolor{revision}{243}}           & \textcolor{revision}{0.44}                                                                       & \textcolor{revision}{4.79}          & \textcolor{revision}{3.11}           & \textcolor{revision}{8.75}          & \multicolumn{1}{c|}{\textcolor{revision}{20.4}}           & \textcolor{revision}{1.29}                                                                       \\
                         & ST                      & 145           & 6.77          & 358           & \multicolumn{1}{c|}{1994}          & 0.41                                                                       & 9.63          & 3.55          & 17.4          & \multicolumn{1}{c|}{31.2}          & 0.41                                                                       \\
                         & STH                     & 22.4          & 5.68          & 101           & \multicolumn{1}{c|}{274}           & 0.81                                                                       & 8.76          & 5.88          & 25.4          & \multicolumn{1}{c|}{32.1}          & 0.81                                                                       \\
                         & \name                   & \textbf{8.57} & \textbf{3.17} & \textbf{16.7} & \multicolumn{1}{c|}{\textbf{29.1}} & 0.68                                                                       & \textbf{3.55} & \textbf{1.69} & \textbf{7.54} & \multicolumn{1}{c|}{\textbf{9.27}} & 1.56                                                                       \\ \hline
\multirow{8}{*}{TW}      & PG                      & 187           & 3.43          & 101           & \multicolumn{1}{c|}{1341}          & -                                                                          & 6.85          & 4.17          & 12.6          & \multicolumn{1}{c|}{25.7}          & -                                                                          \\
                         & Sampling                & 17.3          & \textbf{2.35} & 81            & \multicolumn{1}{c|}{224}           & 0.01                                                                       & 6.03          & 5.32          & 13.1          & \multicolumn{1}{c|}{37.7}          & 0.01                                                                       \\
                         & Greek-S                 & {\ul 12.4}    & 4.47          & {\ul 48.4}    & \multicolumn{1}{c|}{{\ul 97.1}}    & 0.01                                                                       & {\ul 3.22}    & {\ul 2.83}    & {\ul 6.06}    & \multicolumn{1}{c|}{{\ul 10.5}}    & 0.01                                                                       \\
                         & OT-S                    & 19.5          & 3.45          & 109           & \multicolumn{1}{c|}{263}           & 0.06                                                                       & 5.88          & 4.02          & 13.5          & \multicolumn{1}{c|}{31.2}          & 0.06                                                                       \\
                         & Set-Trans                   & \textcolor{revision}{39.5}          & \textcolor{revision}{9.77}           & \textcolor{revision}{108}          & \multicolumn{1}{c|}{\textcolor{revision}{190}}           & \textcolor{revision}{0.81}                                                                       & \textcolor{revision}{4.68}          & \textcolor{revision}{3.06}           & \textcolor{revision}{7.66}          & \multicolumn{1}{c|}{\textcolor{revision}{10.2}}           & \textcolor{revision}{1.54}                                                                       \\
                         & ST                      & 55.7          & 3.37          & 92            & \multicolumn{1}{c|}{768}           & 0.25                                                                       & 7.28          & 4.96          & 14.8          & \multicolumn{1}{c|}{22.3}          & 0.25                                                                       \\
                         & STH                     & 16.7          & {\ul 2.94}    & 64            & \multicolumn{1}{c|}{320}           & 0.47                                                                       & 8.21          & 4.04          & 13.9          & \multicolumn{1}{c|}{30.1}          & 0.47                                                                       \\
                         & \name                   & \textbf{10.1} & 3.22          & \textbf{37.7} & \multicolumn{1}{c|}{\textbf{84.1}} & 0.97                                                                       & \textbf{2.14} & \textbf{1.67} & \textbf{5.28} & \multicolumn{1}{c|}{\textbf{8.22}} & 1.72                                                                       \\ \hline
\end{tabular}%
}
\end{table}

Referring to Table~\ref{tab:dynamic}, \name always has the best performance. Here, we do not report the storage overhead and the estimation latency since they are similar to the ones shown in Tables~\ref{tab:size} and~\ref{tab:latency}. Compared to the static data, PG, Sampling, and Greek-S achieve similar estimation accuracy while the Q-error of other methods increases when working on dynamic data. Compared to Sampling, the update progress of OT-S depends on the trie structure built based on the initial dataset, and this structure does not capture the distribution of elements well when encountering new data. ST and STH incrementally update their corresponding trie structure on dynamic data but fix the generated clusters. However, the latest set always brings a change in the element correlation. Therefore, the elements within the same cluster might be heavily correlated when updating the dataset, leading to performance degradation. In terms of our \name, we also observe performance degradation because the data encoder is trained based on the initial dataset and might not output the best representation of the updated data. However, its performance is more stable than others since we utilize the query information in the query analyzer, which can mitigate this effect.

We also report the average time of each update. Since our \name is both data- and query-driven, it requires training the query analyzer for better performance when the data matrix updates. Compared to \name, all baseline methods do not require any training progress, and thus they need less time to update. However, their representation abilities are not as powerful as that of \name.

\subsection{End-to-End Query Runtime}
\textcolor{revision}{We evaluate the performance of \name in terms of end-to-end query runtime in PostgreSQL. We utilize the latest IMDb dataset~\cite{imdb24dataset} that includes set-valued attributes and extract another database, Food, from the published datasets crawled from \url{http://www.food.com}~\cite{inter2020, review2021, recipes2024}. For the query workload, because current benchmarks, such as JOB~\cite{leis2015good}, do not contain queries with set-valued predicates, we follow the existing work~\cite{meng2023selectivity} to generate queries. Specifically, we firstly use SQLsmith~\cite{sqlsmith2022} and the AI SQL generator~\cite{widenex2024} to generate a query template. Then, we follow the template to generate queries with various granularities. To guarantee the validity of synthetic queries, we follow the same process to generate the set-valued predicates. Note that the set-valued predicate on Food utilize the real user queries, as described in Section~\ref{sec7.3}. Finally, for the IMDB dataset, we generate 30 queries, each containing 4--8 predicates over 3--5 tables, while for the Food dataset, we generate 20 queries, each containing 3--6 predicates over 3 tables.}

To inject cardinalities, our ACE(P) configuration extends the patch from the previous work~\cite{aytimur2024lplm} to accept external estimates for set-valued predicates and the existing support for predicates over categorical and numerical attributes. \textcolor{revision}{In addition to the baseline given by PostgreSQL (PG), we compare with four other baselines. Because ST and STH naturally support queries containing predicates over set-valued attributes, ST and STH configurations inject the estimated cardinalities of ST and STH into the query optimizer for all types of predicates. Note that we use PostgreSQL as the estimator for ST and STH to guarantee a fair comparison. Additionally, configurations GS(P) and Set(P) utilize the same approach as ACE(P) but with estimates from Greek-S and Set-Trans, respectively.}

Table~\ref{tab:e2e} shows the end-to-end (E2E) running time and Q-error. Our method, \name, achieves the best overall performance. Notably, queries with set-valued predicates show a notable improvement due to more accurate estimations. We observe that the E2E time of GS(P) is even longer than that of the PG because the estimation latency of Greek-S is significantly larger.
\begin{table}[htb]
\caption{End-to-end (E2E) time and Q-error}
\vspace{-0.3cm}
\label{tab:e2e}
\resizebox{\columnwidth}{!}{%
\begin{tabular}{|c|ccccc|ccccc|}
\hline
\multirow{2}{*}{Method} & \multicolumn{5}{c|}{IMDB}                                                                                     & \multicolumn{5}{c|}{Food}                                                                                      \\ \cline{2-11} 
                        & Mean & 50\% & 95\% & \multicolumn{1}{c|}{99\%} & \begin{tabular}[c]{@{}c@{}}E2E\\ Time (s)\end{tabular} & Mean & 50\% & 95\% & \multicolumn{1}{c|}{99\%} & \begin{tabular}[c]{@{}c@{}}E2E\\ Time (ms)\end{tabular} \\ \hline
PG                      & 30.3 & 5.11 & 113  & \multicolumn{1}{c|}{440}  & 106.9                                                        & 10.3 & 3.43 & 20.5 & \multicolumn{1}{c|}{79}   & 7327.8                                                        \\
GS(P)                 & 17.4 & 4.66 & 62.1 & \multicolumn{1}{c|}{139}  & 81.1                                                         & 9.41 & 2.78 & 19.2 & \multicolumn{1}{c|}{75.8} & 8007.7                                                        \\
Set(P)               & 19.4 & 4.86 & 80.2 & \multicolumn{1}{c|}{158}  & 82.3                                                         & 8.17 & 2.44 & 16.1 & \multicolumn{1}{c|}{59.1} & 6285.4                                                        \\
ST                      & 12.1 & 3.39 & 44.5 & \multicolumn{1}{c|}{93.1} & 68.2                                                         & 6.53 & 2.05 & 11.5 & \multicolumn{1}{c|}{30.4} & 4667.4                                                        \\
STH                     & 11.8 & 3.47 & 45.1 & \multicolumn{1}{c|}{87.2} & 66.1                                                         & 6.77 & 2.01 & 13.3 & \multicolumn{1}{c|}{30.7} & 4528.6                                                        \\
ACE(P)                  & 10.1 & 2.77 & 33.7 & \multicolumn{1}{c|}{80.1} & 40.5                                                         & 5.53 & 1.78 & 9.01 & \multicolumn{1}{c|}{25.2} & 3004.9                                                        \\ \hline
\end{tabular}%
}
\end{table}

\begin{figure}[tb]
    \centering
    \includegraphics[width=.7\linewidth]{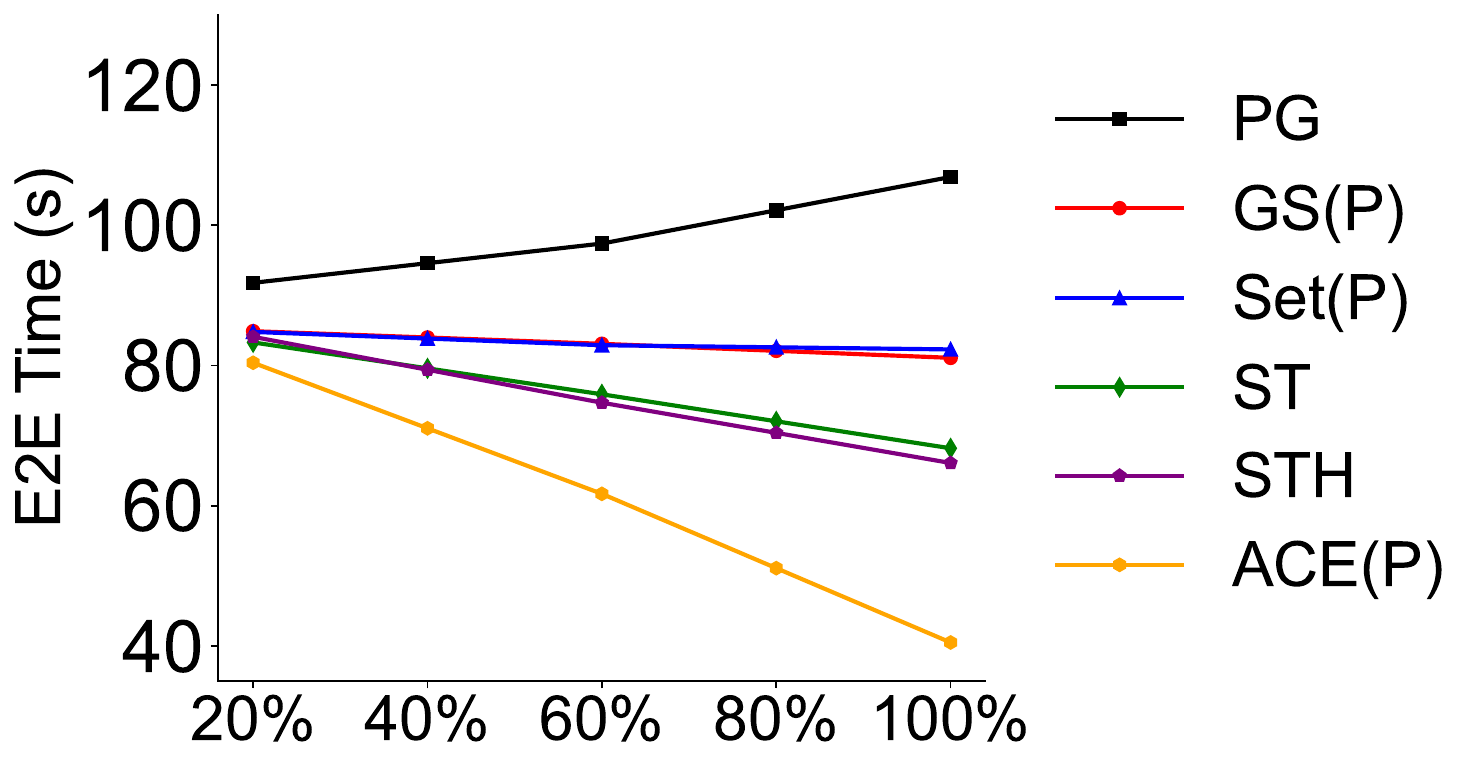}
    \vspace{-0.3cm}
    \caption{E2E time varying the ratio of set-valued queries.}
    \label{fig:e2e-imdb}
\end{figure}
\textcolor{revision}{We also conduct experiments on IMDB to evaluate the E2E performance when varying the number of queries including set-valued predicates (i.e., set queries). Here, we fix the total number of queries in the execution workload and adjust the ratio of set queries. As shown in Figure~\ref{fig:e2e-imdb}, when increasing the ratio, the E2E time of PG increases slightly while the runtime of ACE(P) decreases markedly. We also observe that the results of GS(P) and Set(P) are worse than those of ST and STH because of the worse estimation accuracy or larger estimation latency. The performance for Food is similar and we omit it for brevity.}

\subsection{Ablation Study}
As shown in Table~\ref{tab:abl}, we verify the effectiveness of the main components in \name. Here, we conduct extensive experiments on the WIKI dataset. The results on other datasets are similar and omitted.

\begin{table}[htb]
\caption{Ablation study}
\vspace{-0.3cm}
\label{tab:abl}
\resizebox{\columnwidth}{!}{%
\begin{tabular}{|ccccc|cccc|cccc|cccc|}
\hline
\multicolumn{5}{|c|}{Ablation settings}                                  & \multicolumn{4}{c|}{Subset} & \multicolumn{4}{c|}{Superset} & \multicolumn{4}{c|}{Overlap} \\ \hline
AG           & DS           & CA           & SA           & AP           & Mean  & 50\%  & 95\% & 99\% & Mean  & 50\%  & 95\%  & 99\%  & Mean  & 50\%  & 95\%  & 99\% \\ \hline
$\times$     & $\checkmark$ & $\checkmark$ & $\checkmark$ & $\checkmark$ & 4.47  & 2.69  & 17.2 & 31.8 & 12.6  & 5.34  & 44.4  & 147   & 6.69  & 2.94  & 13.4  & 20.1 \\
$\checkmark$ & $\times$     & $\checkmark$ & $\checkmark$ & $\checkmark$ & 4.11  & 2.72  & 15.6 & 32.1 & 13.7  & 8.41  & 49.9  & 162   & 7.73  & 3.23  & 10.5  & 18.7 \\
$\checkmark$ & $\checkmark$ & $\times$     & $\checkmark$ & $\checkmark$ & 2.87  & 1.58  & 6.26 & 13.7 & 8.19  & 4.14  & 19.8  & 45.8  & 4.31  & 2.45  & 9.74  & 18.3 \\
$\checkmark$ & $\checkmark$ & $\checkmark$ & $\times$     & $\checkmark$ & 3.31  & 1.65  & 5.62 & 11.4 & 8.97  & 3.79  & 25.8  & 55.7  & 5.93  & 3.11  & 11.6  & 22.3 \\
$\checkmark$ & $\checkmark$ & $\checkmark$ & $\checkmark$ & $\times$     & 2.24  & 1.63  & 5.12 & 10.9 & 6.94  & 3.39  & 19.5  & 52.3  & 4.62  & 2.35  & 9.68  & 16.9 \\
$\checkmark$ & $\checkmark$ & $\checkmark$ & $\checkmark$ & $\checkmark$ & 2.04  & 1.36  & 4.93 & 8.71 & 6.44  & 3.34  & 17.9  & 37.4  & 3.98  & 1.75  & 8.99  & 15.8 \\ \hline
\end{tabular}%
}
\end{table}

\noindent \textbf{\uline{Aggregator (AG).}}
To replace our aggregator, we can use traditional methods, such as padding and pooling, to generate the fixed-size set embedding. Since padding often leads to higher storage overhead, we leverage the mean-pooling method, which has a similar cost to our original design. When comparing results, we observe at least a 60\% increase in estimation error at the 50\% quantile. This is because the mean-pooling method typically treats all elements equally, which is not powerful enough to obtain high-quality embeddings.

\noindent \textbf{\uline{Distillation (DS).}}
To replace the distillation module, we propose a random sampling method, setting the sample ratio to 0.001 for a fair comparison. When comparing the results, we observe a significant increase in estimation error ranging from 94.1\% to 112\%. This is because the sampling method captures only a fraction of the dataset's information, whereas the distillation model is designed to compress the matrix while preserving as much information as possible. Additionally, since the sampling method give more attention to high-frequency elements while the low-frequency elements predominantly influences the accuracy of superset queries, we observe the most pronounced fluctuations in these queries.

\noindent \textbf{\uline{Cross-attention (CA).}}
The stacked cross-attention layers serve to link data and queries, mapping the query elements into a latent space to capture their correlation effectively. As the dimensions of the data matrix $\boldsymbol{S_{c}}$ and query element embeddings $\boldsymbol{q_{i}}$ are fixed, we can adopt a straightforward method without the attention mechanism to processing them. For example, 
% for each pair of the data matrix $\boldsymbol{S_{c}}$ and a query element embedding $\boldsymbol{q_{i}}$, 
we flatten $\boldsymbol{S_{c}}$ into a vector and concatenate the vector with $\boldsymbol{q_{i}}$ to generate another vector. Subsequently, the generated vector is fed into a multi-layer perceptron (MLP) with the same number of layers as in our original model. However, experiments reveal that this simplified approach yields worse estimation performance compared to \name, with a decrease exceeding 16.7\%. This performance drop occurs because a straightforward neural network is not powerful enough to discover the implicit relations between elements and data. This finding underscores the necessity and effectiveness of incorporating cross-attention layers.

\noindent \textbf{\uline{Self-attention (SA).}}
The stacked self-attention layers are designed to capture correlations between query elements effectively. To evaluate their contribution, we replace this module with a multi-layer perceptron (MLP). When compared to \name, the modified framework results in 1.49$\times$, 1.31$\times$, and 1.41$\times$ larger Q-error than that of \name at the 99\% quantile for superset, subset, and overlap queries, respectively. These results validate the superiority of self-attention layers in accurately modeling inter-element dependencies, which ultimately leads to more precise cardinality estimation.

\noindent \textbf{\uline{Attention Pooling (AP).}}
The attention pooling module is designed to address the issue of variable-size input while ensuring permutation invaiance. Since any symmetric function can be used to solve this problem, we compare the performance of the attention pooling method with a mean-pooling method. Our analysis verifies the effectiveness of the attention pooling method. Compared to \name, the mean-pooling method results in a slight increase in estimation error, with 1.39$\times$, 1.25$\times$, and 1.07$\times$ larger mean Q-error for superset, subset, and overlap queries, respectively. This performance degradation occurs because mean-pooling weights each embedding equally regardless of its importance~\cite{zafar2022comparison}.
\begin{figure}[tb]
    \centering
    \subcaptionbox{Varying $r$\label{fig:hyper_r}}{
        \vspace{-0.2cm}
        \includegraphics[width=.475\linewidth]{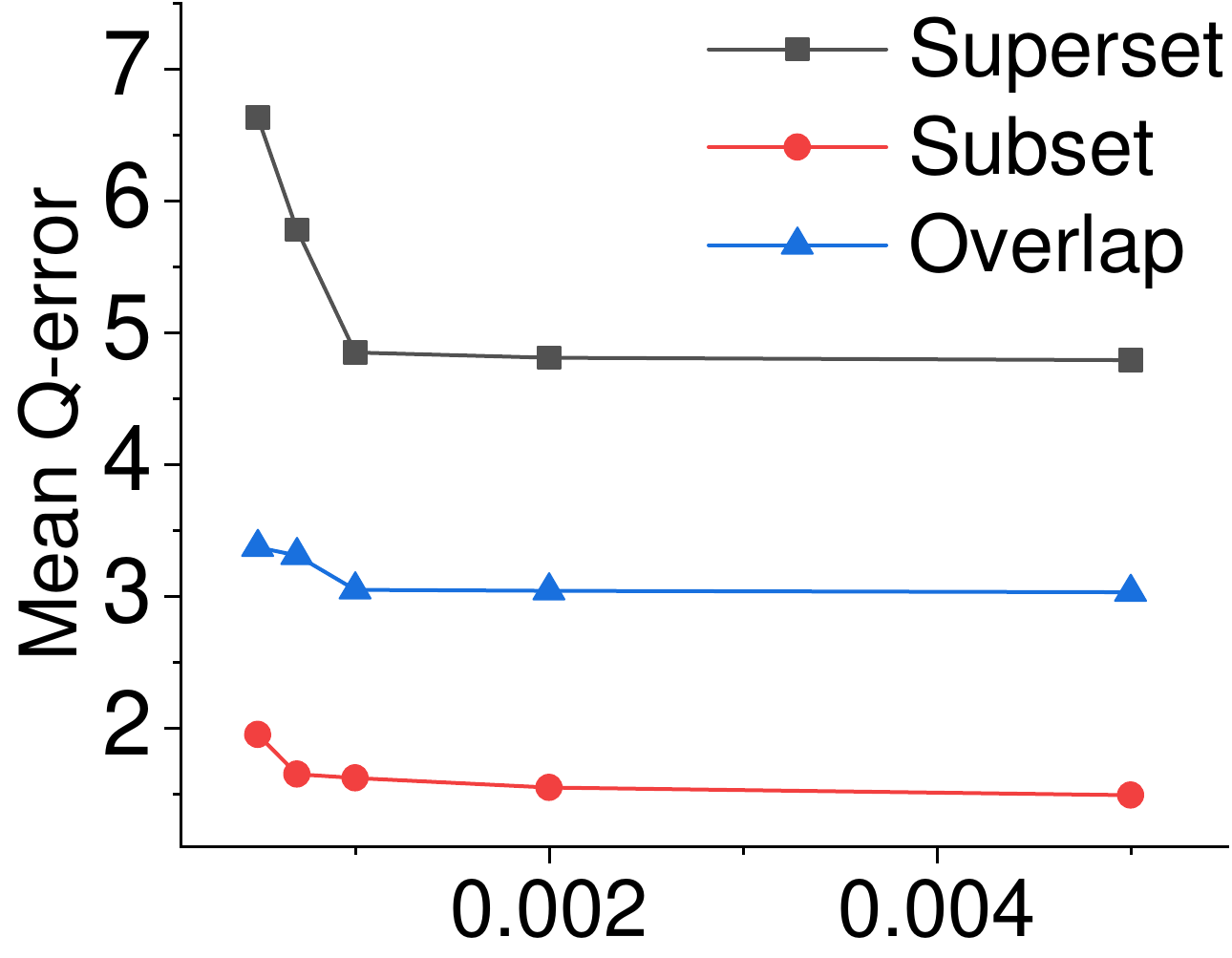}
    }
    \subcaptionbox{Varying $n_{\mathit{distill}}$\label{fig:hyper_dep}}{
        \vspace{-0.2cm}
        \includegraphics[width=.475\linewidth]{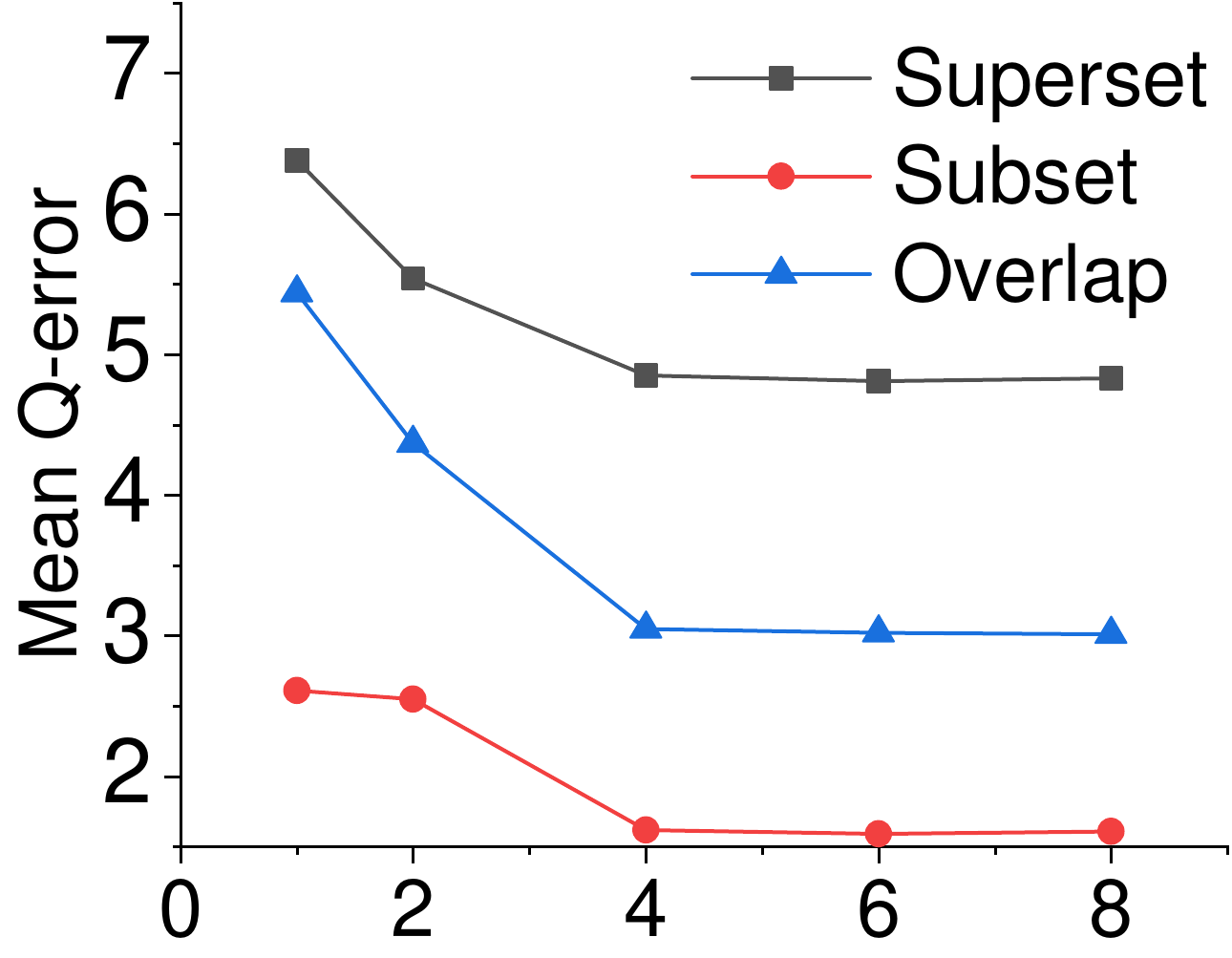}
    }
%     \vspace{-0.3cm}
%     \caption{Estimation performance.}
%     \label{fig:hyper_exp1}
% \end{figure}
% \begin{figure}[tb]
%     \centering
    \subcaptionbox{Varying $n_{\mathit{cross}}$\label{fig:hyper_cross}}{
        \vspace{-0.2cm}
        \includegraphics[width=.475\linewidth]{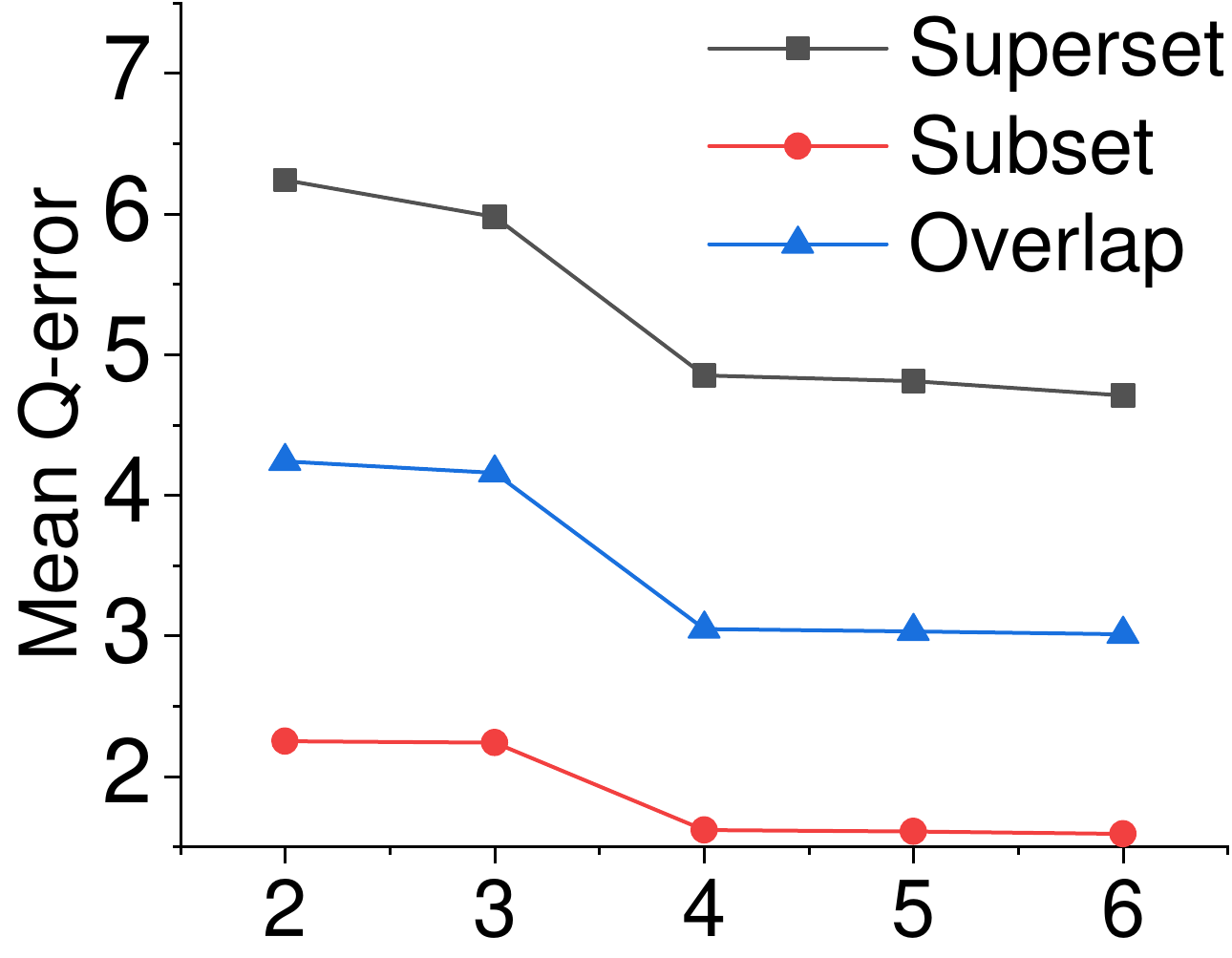}
    }
    \subcaptionbox{Varying $n_{\mathit{self}}$\label{fig:hyper_self}}{
        \vspace{-0.2cm}
        \includegraphics[width=.475\linewidth]{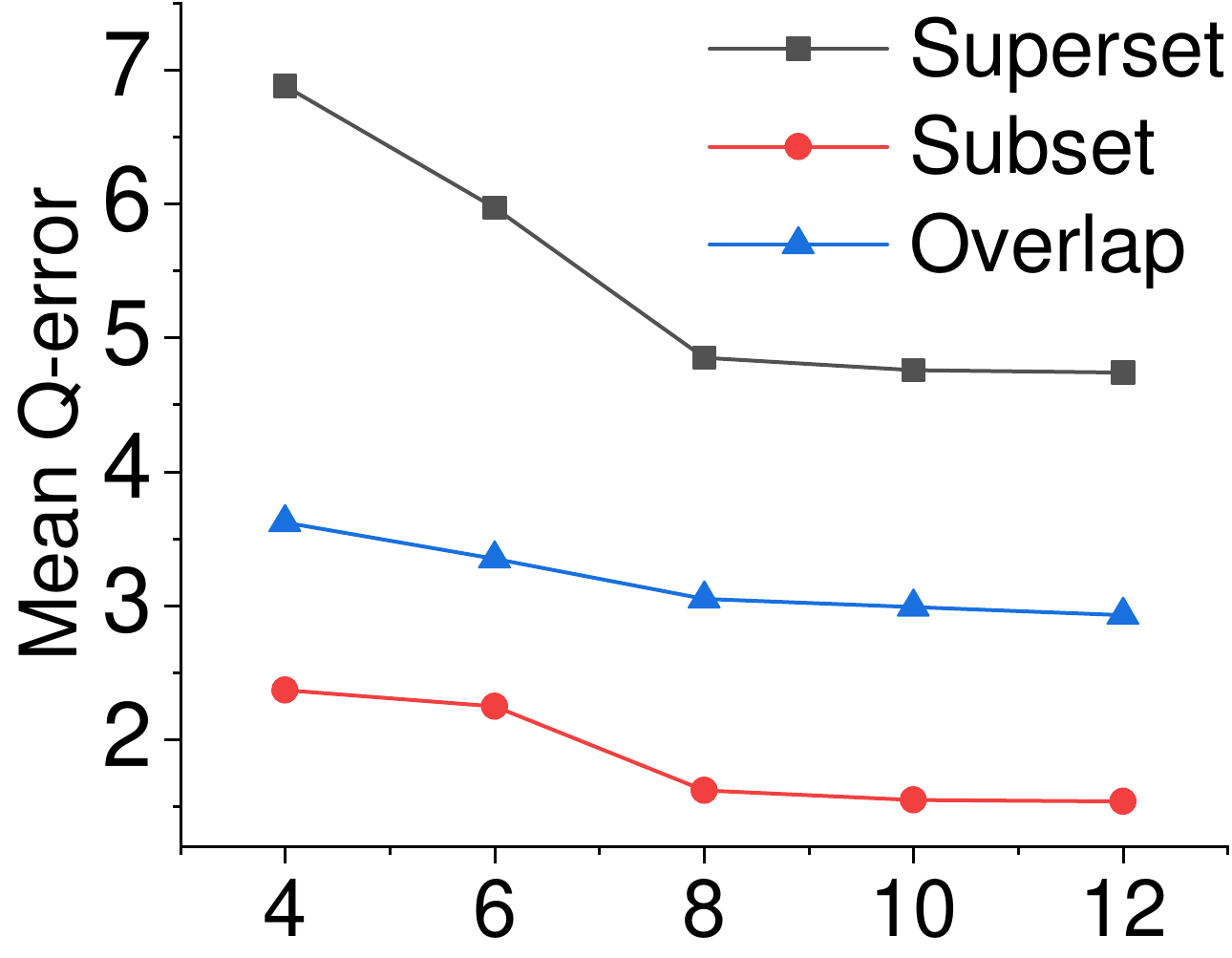}
    }
    \vspace{-0.3cm}
    \caption{Estimation performance.}
    \label{fig:hyper_exp2}
\end{figure}
\begin{figure}[tb]
    \centering
    \subcaptionbox{Varying $r$\label{fig:dis_ratio}}{
        \vspace{-0.2cm}
        \includegraphics[width=.475\linewidth]{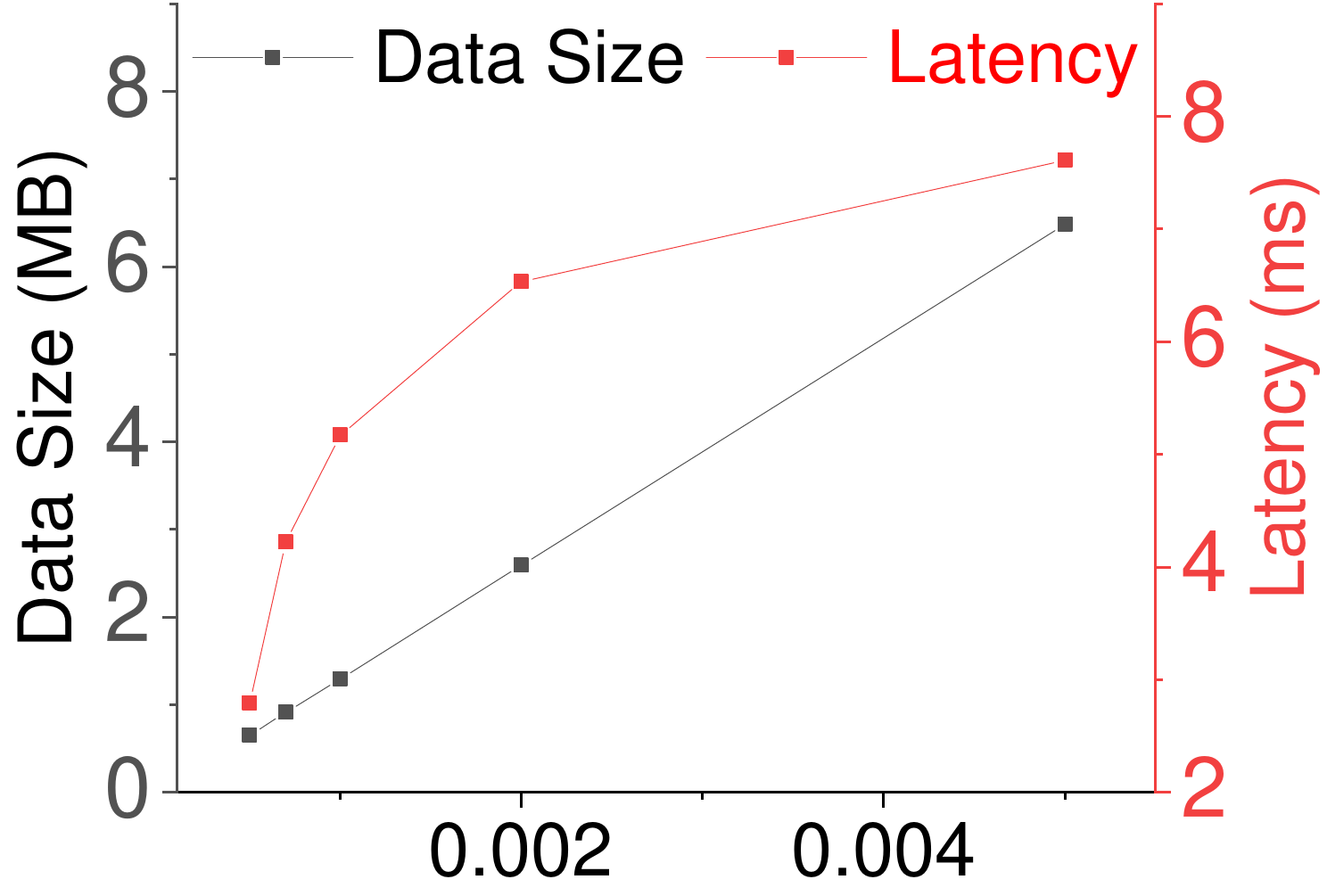}
    }
    \subcaptionbox{Varying $n_{\mathit{distill}}$\label{fig:dis_depth}}{
        \vspace{-0.2cm}
        \includegraphics[width=.475\linewidth]{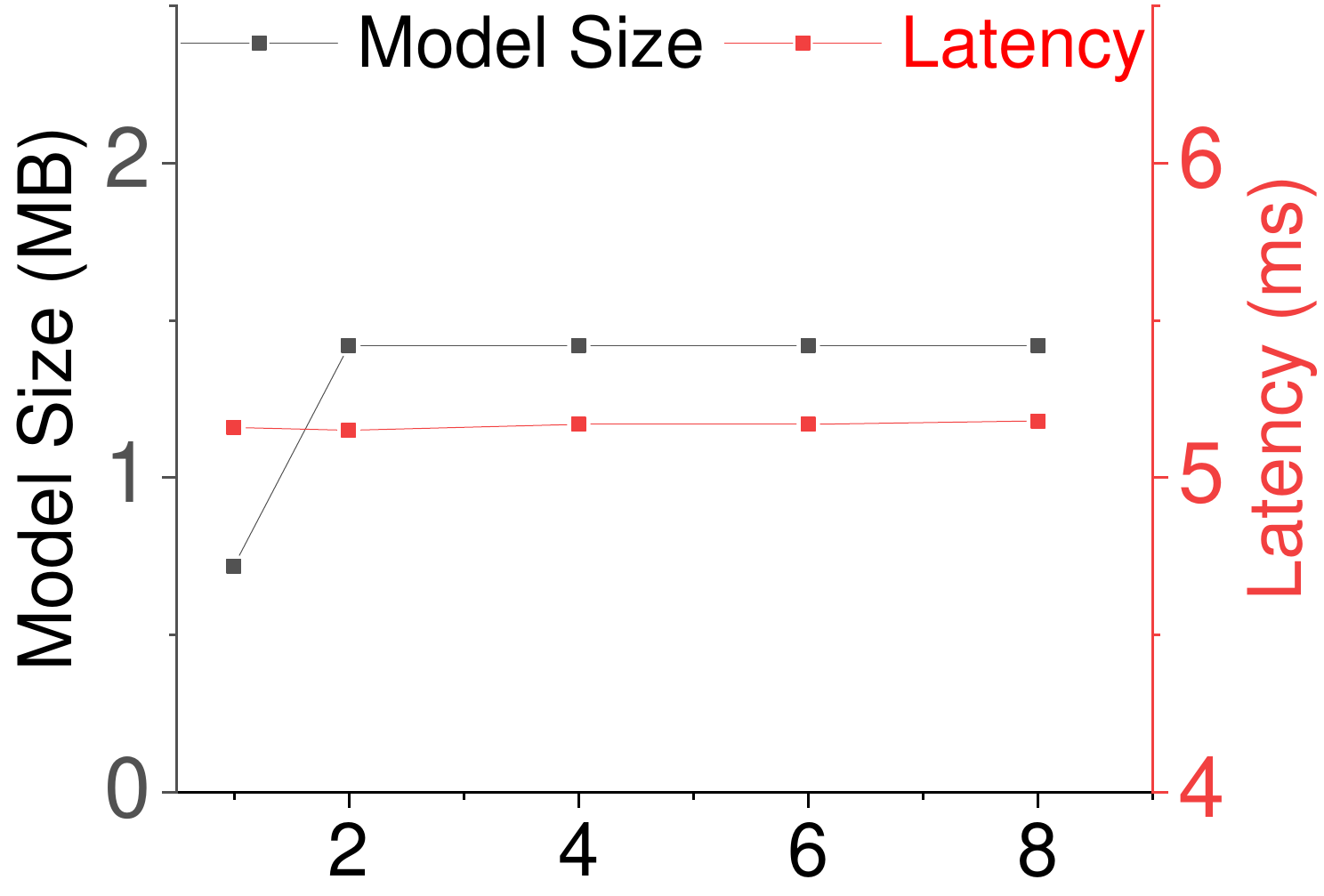}
    }
    \vspace{-0.3cm}
    \caption{Size and latency.}
    \label{fig:dis_exp}
\end{figure}

\subsection{Hyper-parameter Study}
To study the effects of important hyper-parameters, we build different \name versions and observe their performance. Similarly, we only show the comparison results on the WIKI dataset.

\noindent \textbf{\uline{Effects of $r$ and $n_{\mathit{distill}}$.}}
We first study hyper-parameters in our data encoder. Figure~\ref{fig:hyper_r} and~\ref{fig:dis_ratio} show the performance varying distillation ratios $r$. We observe that the size of the distilled matrix is clearly influenced by $r$. When the value of $r$ gets larger, \name produces more accurate estimates, but with higher estimation latency. Besides, the performance improvement becomes marginal when $r$ exceeds 0.001.

Another important hyper-parameter is the number of layers in our distillation model, denoted as $n_{\mathit{distill}}$. Figure~\ref{fig:hyper_dep} illustrates the estimation performance with varying $n_{\mathit{distill}}$. We observe that the Q-error decreases when $n_{\mathit{distill}}$ is less than 4, after which it stabilizes. As shown in Figure~\ref{fig:dis_depth}, the model size remains constant for $n_{distill} \geq 2$ as we share weights between each layer except the first one. Moreover, the estimation latency is similar across these values because the distilled matrices have the same size.

% \begin{figure}[htb]
%     \centering
%     \subcaptionbox{Varying $r$\label{fig:ratio}}{
%         \vspace{-0.2cm}
%         \includegraphics[width=.475\linewidth]{figures/experiments/dis_r.pdf}
%     }
%     \subcaptionbox{Varying $n_{\mathit{distill}}$\label{fig:dis_dep}}{
%         \vspace{-0.2cm}
%         \includegraphics[width=.475\linewidth]{figures/experiments/dis_dep.pdf}
%     }
%     \vspace{-0.3cm}
%     \caption{Distillation performance}
%     \label{fig:mmd}
% \end{figure}

% We also report the distillation performance with varying $r$ and $n_{\mathit{distill}}$ using the maximum mean discrepancy (MMD) values on testing data during the training process. As shown in Figure~\ref{fig:ratio}, a larger $r$ leads to a lower MMD value, indicating a better distilled matrix. Additionally, increasing $n_{\mathit{distill}}$ results in MMD values that eventually become similar, as illustrated in Figure~\ref{fig:dis_dep}. Therefore, we set $r$ and $n_{\mathit{distill}}$ to 0.001 and 4 respectively, which achieve good Q-error performance with acceptable size and estimation latency.

\noindent \textbf{\uline{Effects of $n_{\mathit{cross}}$ and $n_{\mathit{self}}$.}}
We also study the effects of hyper-parameters in our query analyzer. Figures~\ref{fig:hyper_cross} and~\ref{fig:hyper_self} shows the mean Q-error with varying the values of these two hyper-parameters. We observe that increasing $n_{\mathit{cross}}$ and $n_{\mathit{self}}$ both lead to better estimates. This improvement is due to the enhanced ability of more stacked cross-attention layers to discover the relationship between queries and the underlying data, while additional self-attention layers help better capture the correlation between query elements.

\begin{figure}[tb]
    \centering
    \subcaptionbox{Building time}{
        \vspace{-0.2cm}
        \includegraphics[width=.475\linewidth]{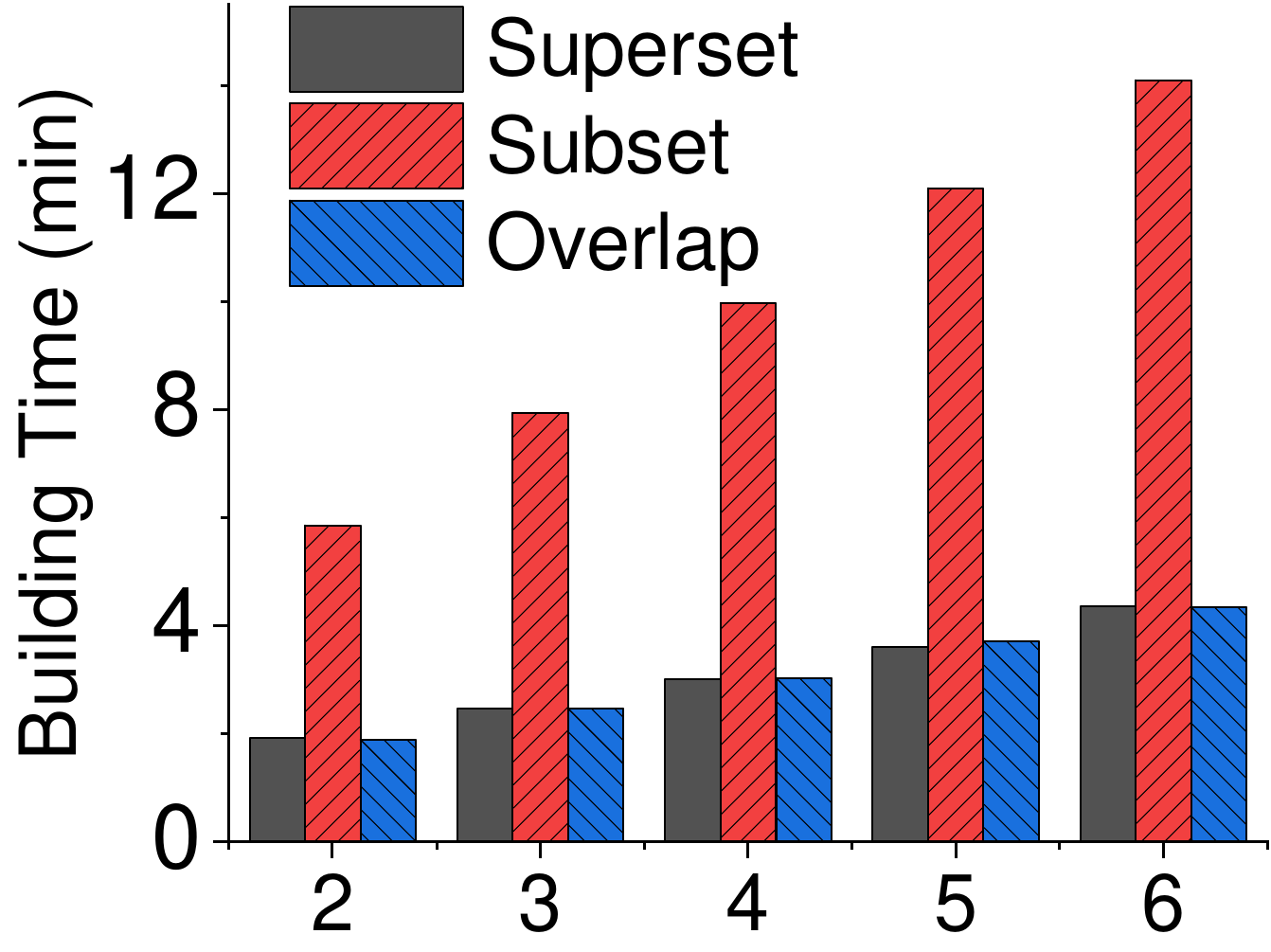}
    }
    \subcaptionbox{Size and latency}{
        \vspace{-0.2cm}
        \includegraphics[width=.475\linewidth]{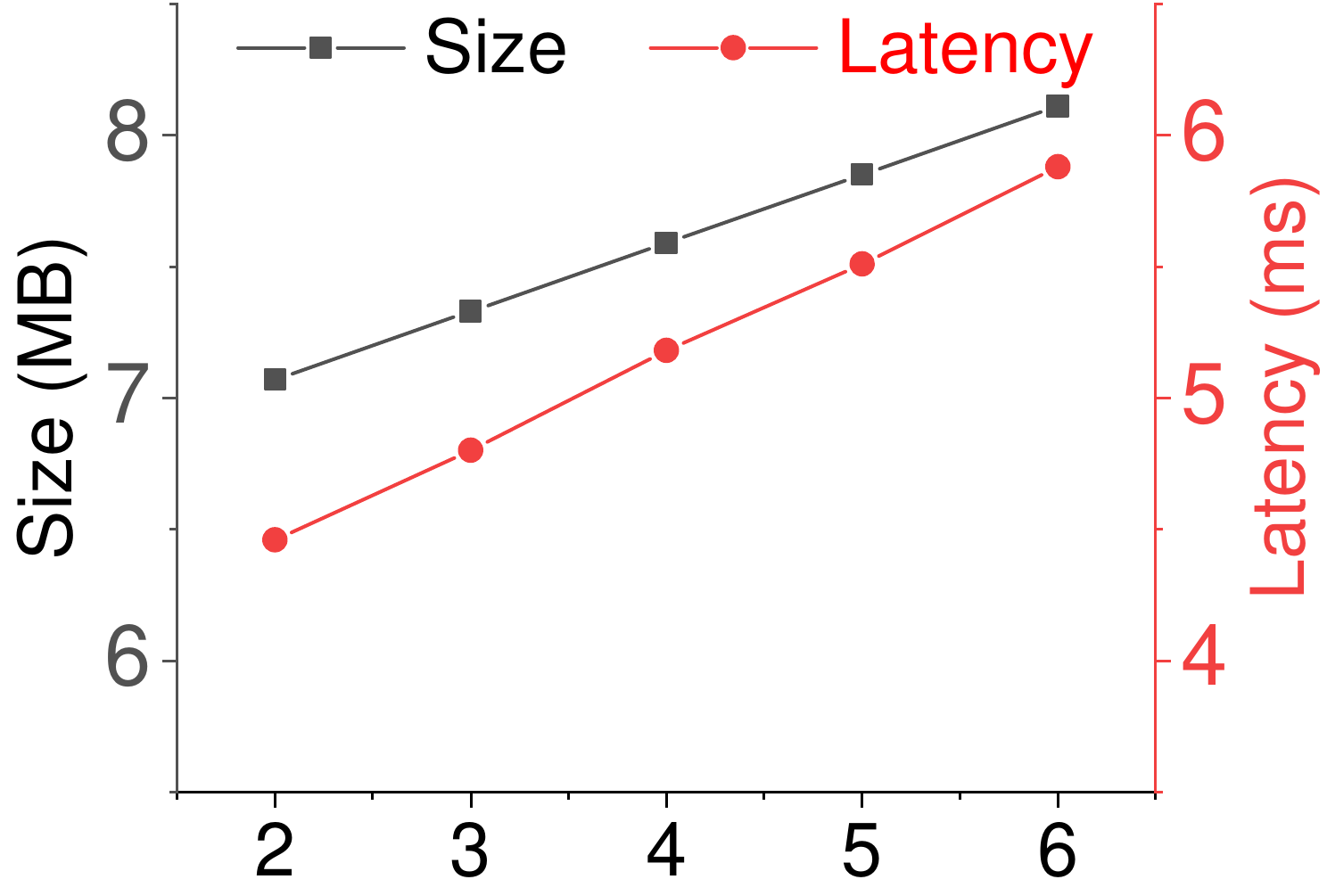}
    }
    \vspace{-0.3cm}
    \caption{Other metrics with varying $n_{\mathit{cross}}$.}
    \label{fig:cross}
\end{figure}
\begin{figure}[tb]
    \centering
    \subcaptionbox{Building time}{
        \vspace{-0.2cm}
        \includegraphics[width=.475\linewidth]{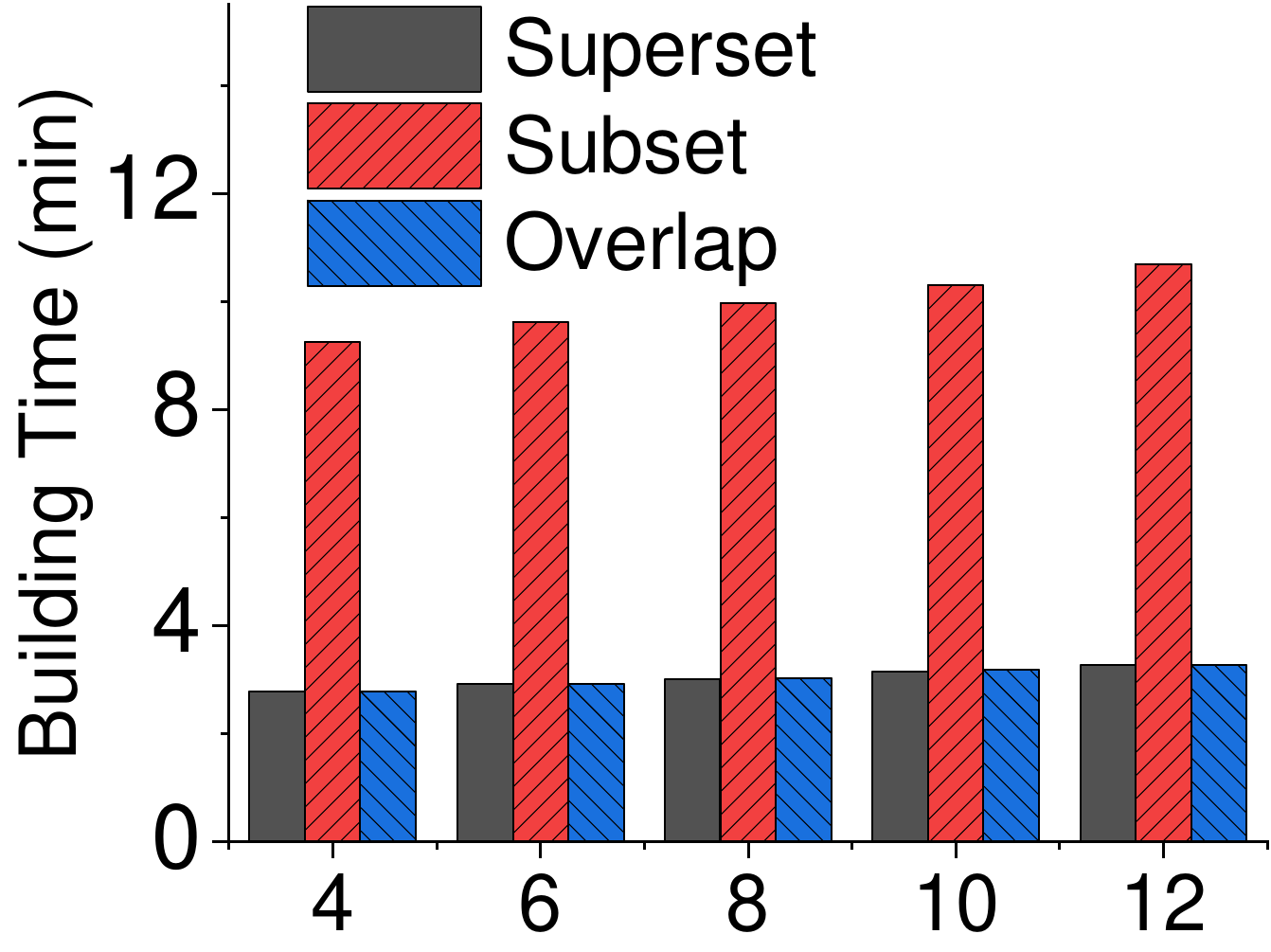}
    }
    \subcaptionbox{Size and latency}{
        \vspace{-0.2cm}
        \includegraphics[width=.475\linewidth]{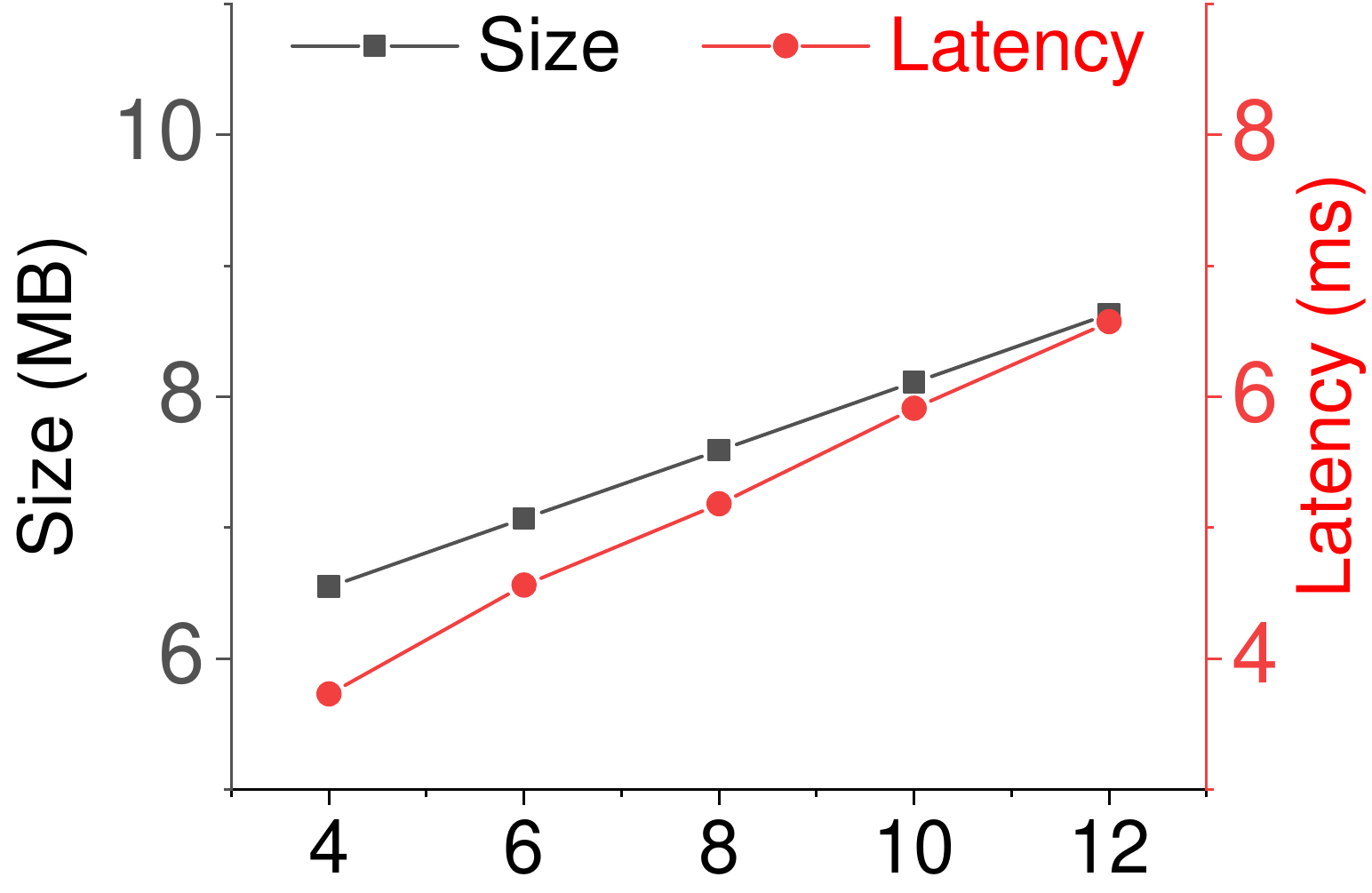}
    }
    \vspace{-0.3cm}
    \caption{Other metrics with varying $n_{\mathit{self}}$.}
    \label{fig:self}
\end{figure}
The values of $n_{\mathit{cross}}$ and $n_{\mathit{self}}$ also affect the building time, the model size, and the estimation latency. Figures~\ref{fig:cross} and~\ref{fig:self} illustrate the experiment results of these metrics. We have two observations. First, larger $n_{\mathit{cross}}$ or $n_{\mathit{self}}$ always leads to a more complex structure, resulting in longer building time, larger model size, and higher estimation latency. Second, the effect of $n_{\mathit{cross}}$ is more significant than that of $n_{\mathit{self}}$ on these metrics. For example, the building time for subset queries increases from about 6 minutes ($n_\mathit{{cross}} = 2$) to 14 minutes ($n_\mathit{{cross}} = 6$), compared to an increase from 9 minutes ($n_\mathit{{self}} = 4$) to 11 minutes ($n_\mathit{{self}} = 12$). This is because the distilled matrix with a larger size is used as one input of the cross-attention sub-module. Therefore, taking into account all aspects, we set the values of $n_{\mathit{cross}}$ and $n_{\mathit{self}}$ to 4 and 8, respectively.

%\vspace{-2ex}
\section{Related Work}
\label{sec8}
\noindent \textbf{Cardinality estimators for numerical and categorical data.} Data-driven methods aim to tightly approximate the data distribution by using statistical or machine learning models. Sampling-based methods~\cite{haas1994relative, lipton1990practical} estimate cardinality from the sampled data. %Their performance relies on the size and quality of sampled data. 
The simple yet efficient 1-D Histogram \cite{selinger1979access} is used in DBMSs such as PostgreSQL. It %assumes that all attributes are independent and 
maintains a histogram for each attribute. %To lift this assumption, 
M-D histogram-based methods \cite{poosala1997selectivity, wang2003multi, muralikrishna1988equi, deshpande2001independence, gunopulos2000approximating} build multi-dimensional histograms to capture the dependency among attributes. However, the decomposition of the joint attributes is still lossy such that they need to make partial independence assumptions. 

Probability models \cite{halford2019approach, tzoumas2011lightweight} utilize the Bayesian network (BN) to model the dependence among attributes, assuming that each attribute is conditionally independent given its parents' distributions. %These models are usually slow in estimation. 
BayesCard \cite{wu2020bayescard} revitalizes BN using probabilistic programming to improve its inference and model construction speed. Deep autoregressive models \cite{hasan2020deep, neurocard, naru} decompose the joint distribution to a product of conditional distributions, which have high accuracy but low efficiency and require large storage space. DeepDB \cite{deepdb} and FLAT \cite{zhu2020flat} build upon a Sum-Product Network (SPN) \cite{poon2011sum} that approximates the joint distribution using multiple SPNs. 

Query-driven methods focus on modeling the relationships between queries and their true cardinalities. LW-XGB and LW-NN~\cite{dutt2019selectivity} formulate the cardinality estimation as a regression problem and apply gradient-boosted trees and neural networks to solve the problem, respectively. The KDE-based join estimator \cite{kiefer2017estimating} combines kernel density estimation (KDE) with a query-driven tuning mechanism. Fauce \cite{liu2021fauce} and NNGP \cite{zhao2022lightweight} assume that the workload follows a Gaussian distribution and adopt deep ensembles \cite{lakshminarayanan2017simple} and neural Gaussian process \cite{lee2018deep} to estimate the mean and variance of the distribution. A few works \cite{wu2021unified, li2023alece} consider both data and workload. These approaches are limited to querying numerical and categorical data, which are difficult to deploy for set-valued data.

\noindent \textbf{Cardinality estimator for set-valued data.} PostgreSQL treats each element as a binary attribute and employs either independence assumptions or the probabilistic model \cite{getoor2001selectivity} to estimate the cardinality of set-valued queries \cite{korotkov2016selectivity}. Yang et al. \cite{yang2019selectivity} improve the sampling method and propose two  estimators: OT-sampling uses a trie structure to focus on highly frequent elements, which struggles with low-frequency elements; DC-sampling leverages the workload type information and employs a divide-and-conquer strategy, which is only applicable for the specified types. 
Hadjieleftheriou et al. \cite{hadjieleftheriou2008hashed} propose a hash sampling algorithm for set similarity queries, which differs from the problem studied in this paper. Meng et al. \cite{meng2023selectivity} propose two algorithms to convert set-valued data into multi-column categorical data and use data-driven methods to estimate the query cardinality. The conversion process can be regarded as approximately solving the NP-hard graph coloring problem, making it difficult to well capture the correlation among elements. All these existing methods only utilize the information of the underlying data. To the best of our knowledge, there is no learning-based estimator that leverages the underlying data and  workload simultaneously.

\noindent \textbf{Attention applications.} %Due to the prevalence of Transformer~\cite{vaswani2017attention}, 
The attention mechanism has been applied to various problems~\cite{wang2017residual,zheng2020gman,xu2024timesgn,zhang2022mderank,xu2025fast}. Recently, it has been adapted for database optimization~\cite{ge2021watuning}. The most closely related work~\cite{li2023alece} estimates the cardinality for SPJ (Select-Project-Join) queries. However, its featurization method is not suitable for our problem as the number of columns (less than 100) is significantly smaller than the number of sets (exceeding $10^6$). Thus, our \name needs new designs for the data encoder and the query analyzer.

\section{Conclusion}% and Future Work}
\label{sec9}
We presented \name, a versatile learned cardinality estimation model that makes high-quality estimates for set-valued queries. We first propose a distillation-based data encoder to represent the entire dataset using a compact matrix. To capture correlations between query elements, we then propose an attention-based query analyzer. Since query lengths can vary, we employ a pooling module to derive the fixed-size vector. Extensive experimental results demonstrate superior performance of \name compared to the state of the art.

%For future work, it is interesting to extend \name to support more query types, such as categorical, numerical, and even string predicates. Also, exploring better distillation methods that can further reduce storage overhead is a relevant direction.

% \clearpage
\balance
\bibliographystyle{ACM-Reference-Format}
\bibliography{ref-scholar}

\appendix
\section{More Experiments}
\subsection{Details of Queries}
We analyze the overlap ratio of low-frequency elements in the test and training data. Table~\ref{tab:low-ratio} shows the result on the WIKI dataset, which includes the number of overlapped low-frequency elements and the number of distinct low-frequency elements that appeared in the training data, respectively. It can be observed that the model can only see a very small portion of the low-frequency elements in the test data when training. Other datasets have similar observations.
\begin{table}[htb]
\caption{Low-frequency element overlap ratio}
\vspace{-0.3cm}
\label{tab:low-ratio}
\begin{tabular}{|c|c|c|}
\hline
Subset    & Superset & Overlap \\ \hline
151/11077 (1.3\%) & 13/651 (1.9\%)  & 12/805 (1.5\%) \\ \hline
\end{tabular}
\end{table}

\subsection{A Decomposition Baseline}
A na\"ive baseline method is to replace the set columns as a table (essentially representing the bipartite graph) and replacing the set based query with a regular query involving this new table and applying regular cardinality estimation methods. However, this method has shown worse performance than one of our baselines, PG, in the former study~\cite{korotkov2016selectivity}. We omit it due to space limit. To verify this result, we conduct experiments on the WIKI dataset. In Table~\ref{tab:dtpg}, DT and PG denote the decomposition method and the PG baseline in our paper, respectively. It can be observed that PG outperforms DT in most cases.
\begin{table}[htb]
\caption{Comparison between the decomposed table (DT) and PG}
\vspace{-0.3cm}
\label{tab:dtpg}
\resizebox{\columnwidth}{!}{%
\begin{tabular}{|c|cccc|cccc|cccc|}
\hline
\multirow{2}{*}{Method} & \multicolumn{4}{c|}{Subset} & \multicolumn{4}{c|}{Superset} & \multicolumn{4}{c|}{Overlap} \\ \cline{2-13} 
                        & Mean  & 50\%  & 95\% & 99\% & Mean  & 50\%  & 95\%  & 99\%  & Mean  & 50\%  & 95\%  & 99\% \\ \hline
DT                      & 22.5  & 5.98  & 106  & 199  & 294   & 33.4  & 584   & 2362  & 8.41  & 1.72  & 10.3  & 105  \\
PG                      & 9.49  & 5.68  & 19.7 & 41.3 & 175   & 8     & 271   & 1785  & 8.28  & 1.81  & 13.1  & 94.2 \\ \hline
\end{tabular}%
}
\end{table}

Although the decomposition method can be lossless, there are two major drawbacks. First, its storage overhead is extremely large. For example, the decomposed table occupies 2,715 MB, 5$\times$ larger than the original table. Second, set-valued queries need to be rewritten, making them much more complex than the original ones. One example of the superset query is \texttt{SELECT * FROM \textit{parent\_table} p WHERE NOT EXISTS (SELECT * FROM \textit{child\_table} c WHERE p.id = c. parent\_id AND c.value NOT IN (val\_1, val\_2, ..., val\_n);}

Set-valued data has emerged as an essential data type in various applications to model the one-to-many relationship and to store mulit-valued attributes. Our proposed estimator aims to obtain accurate cardinality estimation results for queries containing set-valued predicates thus generating more optimized query execution plans.

\subsection{The Study of Element Frequency}
\begin{figure}[htb]
   \centering
   \subcaptionbox{Subset predicate\label{fig:e2e_sub}}{
       \vspace{-0.2cm}
       \includegraphics[width=.35\linewidth]{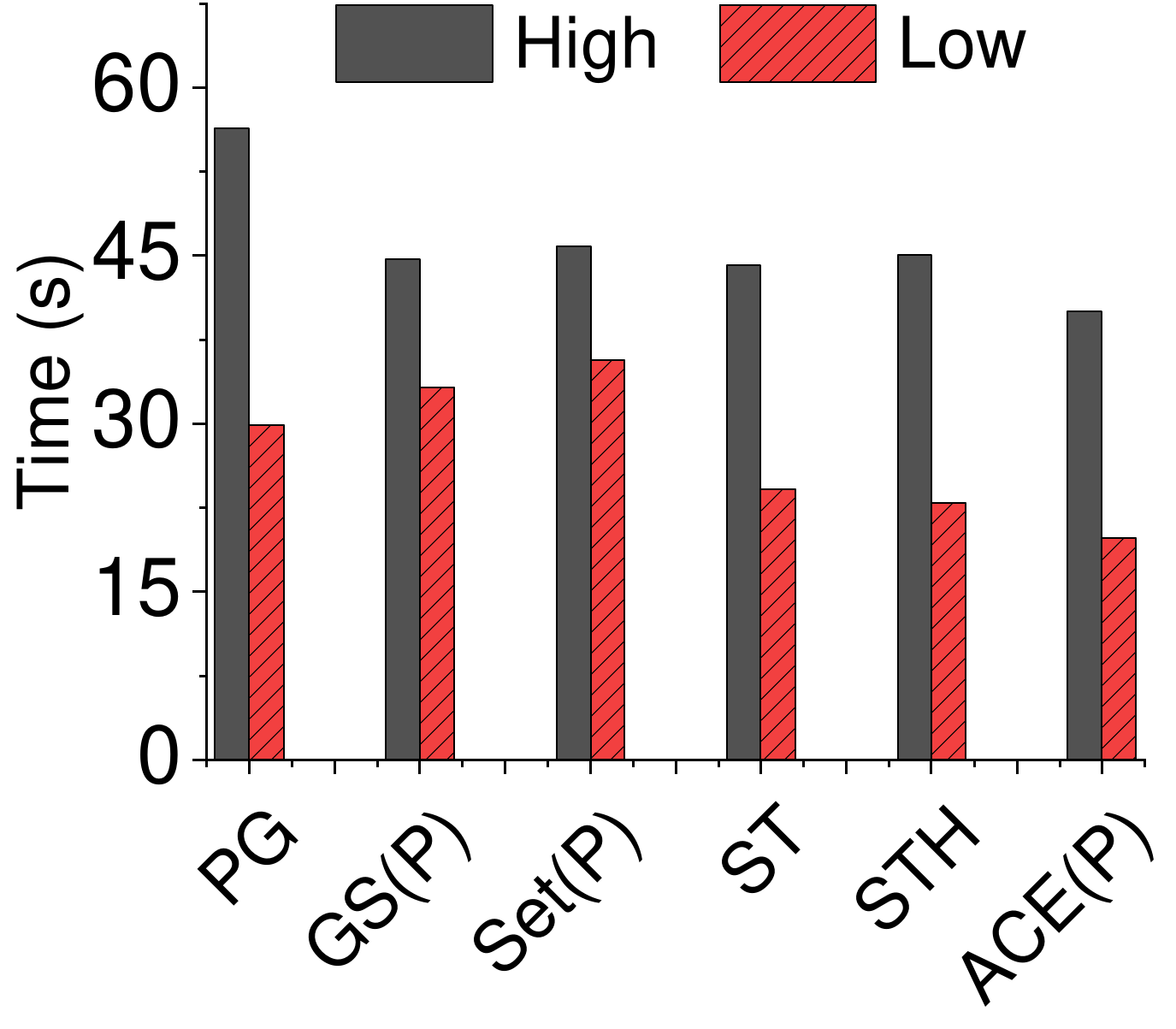}
   }
   \subcaptionbox{Superset predicate\label{fig:e2e_sup}}{
       \vspace{-0.2cm}
       \includegraphics[width=.35\linewidth]{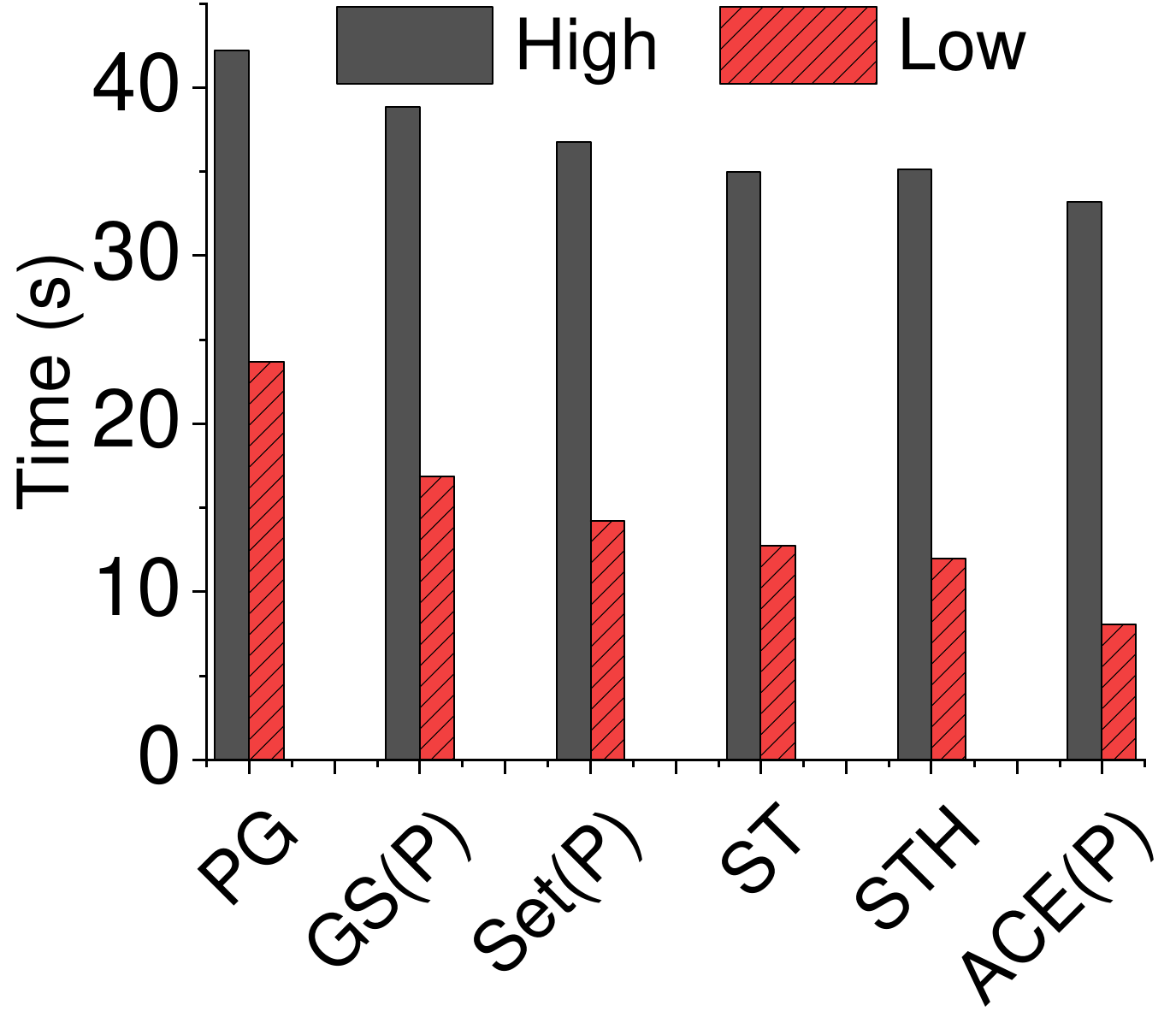}
   }
   \vspace{-0.3cm}
   \caption{End-to-end query runtime.}
   \label{fig:e2e}
\end{figure}

We conduct experiments on the effect of the query elements' frequencies. We utilize the generation process mentioned in Section~\ref{sec7.1} and obtain two collections of queries with only high-frequency and low-frequency predicates, respectively. 
As shown in Figure~\ref{fig:e2e}, \name performs better on queries with low-frequency predicates because a more accurate estimation result significantly affects the execution plan in such a case. Another interesting thing is that the E2E time of GS(P) and Set(P) on queries with low-frequency subset predicates is even greater than PG because a larger estimation error affects the executed query plan.

% Figure~\ref{fig:e2e} shows the performance of all methods. We observe that the improvement on low-frequency predicate is more significant than high-frequency ones because of more accurate estimates, leading to the change in the execution plan of most queries with the low-frequency predicate. Since ST and STH utilize SOTA estimators, which can capture the correlation between attributes, their performance is better than GS(P) and Set(P). Another interesting thing is that the E2E time of GS(P) and Set(P) on queries with low-frequency subset predicates is even larger than PG because larger estimation error affects the executed query plan.

\end{document}